\documentclass[twocolumn, twocolappendix]{aastex631}

\usepackage{pifont} 
\usepackage{tikz}
\usepackage{hyperref}

\usepackage{xcolor}
\definecolor{red}{rgb}{0.96, 0.36, 0.36}

\begin{document}

\definecolor{pot1}{HTML}{AE76A3} 
\definecolor{pot2}{HTML}{1965B0} 
\definecolor{pot3}{HTML}{7BAFDE} 
\definecolor{pot4}{HTML}{4EB265} 
\definecolor{pot5}{HTML}{F7F056} 
\definecolor{pot6}{HTML}{E8601C} 
\definecolor{pot7}{HTML}{DC050C} 
\definecolor{pot8}{HTML}{72190E} 

\title{LMC Calls, Milky Way Halo Answers: Disentangling the Effects of the MW--LMC Interaction on Stellar Stream Populations}

\correspondingauthor{Richard A. N. Brooks}
\email{richard.brooks.22@ucl.ac.uk}

\author[0000-0001-5550-2057]{Richard A. N. Brooks}
\affiliation{Department of Physics and Astronomy, University College London, London, WC1E 6BT, UK}\affiliation{Center for Computational Astrophysics, Flatiron Institute, Simons Foundation, 162 Fifth Avenue, New York, NY 10010, USA}

\author[0000-0001-7107-1744]{Nicolás Garavito-Camargo}
\affiliation{Center for Computational Astrophysics, Flatiron Institute, Simons Foundation, 162 Fifth Avenue, New York, NY 10010, USA}

\author[0000-0001-6244-6727]{Kathryn V. Johnston}
\affiliation{Department of Astronomy, Columbia University, New York, NY 10027, USA}

\author[0000-0003-0872-7098]{Adrian~M.~Price-Whelan}
\affiliation{Center for Computational Astrophysics, Flatiron Institute, Simons Foundation, 162 Fifth Avenue, New York, NY 10010, USA}

\author[0000-0003-4593-6788]{Jason L. Sanders}
\affiliation{Department of Physics and Astronomy, University College London, London, WC1E 6BT, UK}

\author[0000-0001-9046-691X]{Sophia Lilleengen}
\affiliation{Institute for Computational Cosmology, Department of Physics, Durham University, South Road, Durham DH1 3LE, UK}



\begin{abstract}

The infall of the LMC into the Milky Way (MW) has dynamical implications throughout the MW's dark matter halo.
We study the impact of this merger on the statistical properties of populations of simulated stellar streams. 
Specifically, we investigate the radial and on-sky angular dependence of stream perturbations caused by the direct effect of stream--LMC interactions and/or the response of the MW dark matter halo. 
We use a time-evolving MW--LMC simulation described by basis function expansions to simulate streams.
We quantify the degree of perturbation using a set of stream property statistics including the misalignment of proper motions with the stream track.
In the outer halo, direct stream--LMC interactions produce a statistically significant effect, boosting the fraction of misaligned proper motions by $\sim 25\%$ relative to the model with no LMC. 
Moreover, there is on-sky angular dependence of stream perturbations:~the highest fractions of perturbed streams coincide with the same on-sky quadrant as the present-day LMC location.
In the inner halo, the MW halo dipole response primarily drives stream perturbations, but it remains uncertain whether this is a detectable signature distinct from the LMC's influence.
For the fiducial MW--LMC model, we find agreement between the predicted fraction of streams with significantly misaligned proper motions, $\bar{\vartheta}>10^{\circ}$, and \textit{Dark Energy Survey} data.
Finally, we predict this fraction for the Rubin Observatory \textit{Legacy Survey of Space and Time} (LSST) footprint. Using LSST data will improve our constraints on dark matter models and LMC properties as it is sensitive to both.

\end{abstract}



\section{Introduction} \label{sec:intro}


Our Galaxy, the Milky Way (MW) is undergoing a merger with the LMC\footnote{The Large Milky/Magellanic Cloud. See \citet{Vasiliev2023} for a comprehensive review detailing the effect of the LMC on the MW.}. The LMC is thought to be on its first pericentric passage and to have a dark matter mass $M_{\mathrm{LMC}}\sim 10^{11}\,\mathrm{M}_{\odot}$\citep{Besla2007, Besla2010, Boylan-Kolchin2011, Penarrubia2016, Kravtsov2024}. An alternative scenario has the LMC on its second pericentric passage \citep{Vasiliev2024}. Though, most features of the earlier passage are superseded by the more recent passage at a smaller pericenter. 
Many Local Group phenomena require a large mass for the LMC to explain: for example, the kinematics of MW satellites \citep{CorreaMagnus2022, Kravtsov2024}; dynamical models of stellar streams \citep{2019MNRAS.487.2685E, Koposov2019, Shipp2021, Vasiliev2021a}; and the timing argument (\citealt{Penarrubia2016}, but see also \citealt{Benisty:2022, Chamberlain:2023, Benisty2024}). The LMC has also been observed to generate significant disequilibrium in the MW gravitational potential:~the displacement of the MW disc \citep{2020MNRAS.494L..11P, Vasiliev2021a}, a stellar over-density \citep{Belokurov2019, Garavito-Camargo2019, Conroy2021, Amarante2024}, and the reflex motion of the stellar halo \citep{2019MNRAS.487.2685E, 2020MNRAS.494L..11P, Petersen2021, Erkal2021, Yaaqib2024, Chandra2024}. The orbit of the LMC is sensitive to the assumed Galactic potential \citep[see fig.~3 of][]{Vasiliev2023} and, because the LMC is of considerable mass, it is also subject to dynamical friction from the MW dark matter halo \citep{Chandrasekhar1943}. Current state-of-the-art models of the MW--LMC system account for dynamical friction and the reflex motion of both galaxies \citep[e.g.,][]{Gomez2015, Patel2017, Erkal2019, Garavito-Camargo2019, Patel2020, Cunningham2020, Vasiliev2021a, Garavito-Camargo2021, Dillamore2022, Donaldson2022, Koposov2023, Lilleengen2023, Vasiliev2024}.

Stellar streams form when satellites, dwarf galaxies or globular clusters orbiting the MW tidally disrupt. Stellar streams are kinematic tracers of the Galactic potential \citep{Johnston1999, Helmi1999} as stars within a stream roughly delineate orbits in the host potential \citep{McGlynn1990, Johnston1996, Binney2008, Sanders2013}. This allows us to infer the accelerations that the stars experience and hence the host’s gravitational field. In the MW, we know of $\sim 170$ globular clusters \citep{Vasiliev2021b} and $\sim 80$ MW stellar streams\footnote{See \citet{Bonaca2024} for a review of MW streams in the \textit{Gaia} era \citep{GaiaCollaboration2021}.} associated with globular clusters or coming from fully disrupted globular clusters \citep{Mateu2023}. Most of the detected globular cluster streams are within a Galactocentric distance of $\lesssim 35\,\mathrm{kpc}$. However, the current number of known MW streams \citep{Mateu2023} is likely not the full population in the MW \citep{Shipp2023, Pearson2024} and we can expect to discover many more with future surveys, e.g. Rubin Observatory \textit{Legacy Survey of Space and Time}  \citep[LSST,][]{Ivezic2019}, out to much greater Galactocentric distances of $\sim 75\,\mathrm{kpc}$. 

At a Galactocentric distance of $\sim 50\,\mathrm{kpc}$ \citep{Pietrzynski2019}, the LMC has perturbed streams in the MW, especially those with close encounters \citep[e.g. Orphan-Chenab,][]{2019MNRAS.487.2685E, Shipp2021, Lilleengen2023, Koposov2023, Brooks2024}. As we expect to uncover more outer halo MW streams \citep{Pearson2024}, it is useful to predict the extent to which the LMC has left them dynamically perturbed. Testing these predictions will allow us to learn the properties of the LMC itself. Additionally, the Galactic suburbs are ideal for studying deformations of streams caused by interactions with dark matter subhaloes from both the MW and LMC \citep[e.g.,][]{Yoon2011, Sanders2016, Erkal2017, Bonaca2019, Bonaca2020, Tavangar2022, Hilmi2024, Bonaca2024} because of fewer additional disturbances from dynamical resonances and baryonic interactions in the inner parts of our Galaxy \citep[e.g.,][]{Price-Whelan2016, Pearson2017, Kawata2021}.
In fact, the frequency of subhalo--stream interactions and the resulting perturbed stream properties are dependent on the underlying cosmological model \citep[e.g. cold vs warm dark matter,][]{Carlberg2024b, Carlberg2024a}. 
Therefore, it is of great importance for us to further disentangle perturbations due to small scale structure, i.e., dark matter subhaloes, and/or bulk effects such as the LMC.

As the LMC, a satellite galaxy, falls into the gravitational potential of the MW, the central galaxy, the host responds by generating a density wake \citep{Chandrasekhar1943}. This is because an infalling satellite will have a broad range of orbital frequencies that resonate with dark matter particles of the host galaxy \citep{Mulder1983, Weinberg1986}. 
The classical `conic' wake trailing the LMC is described as the \textit{transient response}, whereas the response elsewhere in the MW halo is the \textit{collective response} caused by the amplitude of the barycenter displacement \citep{Garavito-Camargo2019, 2021ApJ...919..109G, Tamfal2021, Foote2023}. 
The detailed structure of these deformations depends on the nature of dark matter itself \citep[e.g.,][]{Furlanetto2002, Hui2017, Lancaster2020}.
The density wake of the LMC is predicted to leave an observable signature in the density and kinematics of MW halo stars \citep[e.g.,][]{Conroy2021, Cavieres2024}, including those in stellar streams. Any perturbations due to the wake will be most striking in the outer halo trailing the past orbit of the LMC, $\gtrsim 50\,\mathrm{kpc}$ \citep[e.g.,][]{Vasiliev2023}. However, there are very few streams that have been detected at, or beyond, these distances to date. 

\setlength{\tabcolsep}{3pt}
\begin{table*}
\centering
\caption{Summary table for the MW \& LMC N-body models, Sec.~\ref{sec:methods-Nbody}. Virial quantities are defined by a sphere enclosing an overdensity that is 200 times the critical density of the Universe, $\rho_{\mathrm{crit}} = 3H^{2}/8\pi G$, denoted with a ‘vir’ subscript.}
\begin{tabular}{ccccc} 
\hline
\hline
& \shortstack{\\ MW dark\\ matter halo} & \shortstack{\\ MW stellar\\ disc} & \shortstack{\\ MW stellar\\ bulge}  & \shortstack{\\ LMC dark\\ matter halo}\\
\hline
\hline
 \shortstack{\\Potential \\profile} & \shortstack{\\Navarro-Frenk-White \\\citep{Navarro1996, Navarro1997}} & \shortstack{\\Miyamoto-Nagai \\\citep{Miyamoto1975}} & \shortstack{\\Hernquist \\\citep{Hernquist1990}} & \shortstack{\\Hernquist \\\citep{Hernquist1990}} \\
\shortstack{\\Potential \\parameters} & \shortstack{\\$M_{\mathrm{vir}} = 7.92\times10^{11}\,\mathrm{M}_{\odot}$ \\ $r_{s} = 12.8\,\mathrm{kpc}$ \\ $c=15.3$} & \shortstack{\\$M_{\mathrm{disc}} = 6.8\times10^{10}\,\mathrm{M}_{\odot}$ \\ $a = 3.0\,\mathrm{kpc}$ \\ $b = 0.28\,\mathrm{kpc}$} & \shortstack{$M_{\mathrm{bulge}} = 5.0\times10^{9}\,\mathrm{M}_{\odot}$ \\ $r_s = 0.5\,\mathrm{kpc}$} & \shortstack{$M_{\mathrm{LMC}} = 1.25\times10^{11}\,\mathrm{M}_{\odot}$ \\ $r_{s} = 14.9\,\mathrm{kpc}$}\\
\shortstack{\\$N$-body\\particles} & $10^7$ & $10^6$ & $10^6$ & $10^7$ \\
$l_{\mathrm{max}}$ & 6 & -- & -- & 6 \\
$m_{\mathrm{max}}$ & -- & 6 & 6 & -- \\
$n_{\mathrm{max}}$ & 17 & 17 & 17 & 23 \\
\hline
 \label{table1}
\end{tabular}
\end{table*}

Basis function expansions (BFEs) are used to represent general mass distributions as linear combinations of basis functions. As such, BFEs offer the flexibility to model the deformations captured in $N$-body simulations \citep{Lilley2018a, Lilley2018b, Sanders2020, 2020MNRAS.494L..11P, 2021ApJ...919..109G, Lilleengen2023}. In this work, we use the $N$-body simulation of \citet{Lilleengen2023} which utilise a BFE description using the \textsc{exp} toolkit \citep{2022MNRAS.510.6201P}. This provides a time-evolving MW system in which stellar streams can be generated. These simulations account for the deformations to the MW and LMC dark matter haloes since the latter's infall, including the formation of a dark matter dynamical friction wake trailing the LMC. Studies of $N$-body MW--LMC simulations act as complementary efforts to MW--LMC analogues identified from cosmological simulations \citep[e.g., the \textit{Milky Way-est} cold dark matter, and DREAMS warm dark matter, zoom in simulations,][respectively]{Buch2024, Rose2024}.

Perturbations to streams caused by interactions with the LMC have been studied for a handful of known MW streams e.g., the LMC has been attributed to the misalignment of proper motions in the Orphan-Chenab stream \citep{Erkal2019}, the Sagittarius stream \citep{Vasiliev2021a} and for the \textit{Dark Energy Survey} \citep[DES,][]{Shipp2018, Shipp2019} streams, with the stream perturbations being a sensitive tracer of the properties of the LMC \citep[e.g its mass,][]{Erkal2019, Shipp2021, Koposov2023}. In this work, we will statistically determine the extent of two distinct effects on simulated populations of realistic MW stellar streams. Specifically, we will investigate the radial and on-sky angular dependence of stream perturbations caused by:~direct interactions with the LMC and/or the MW halo deforming in response to the infalling LMC. We will compare predictions to the current observational data from DES and make predictions for what LSST can expect to observe. These predictions are sensitive to:~(i) the underlying dark matter model e.g., varying the dark matter model on galaxy cluster scales alters the galaxy-dark matter offset, density profile and subhalo distribution \citep[][]{Kim2017, Banerjee2020}. As gravity is scale-invariant, similar behaviour is expected in the MW \citep[e.g.,][]{Sameie2018}. And (ii) the properties of the LMC e.g., a more/less massive LMC would increase/decrease the fraction of streams displaying significant perturbations.
Hence, future observations in conjunction with flexible models of the MW--LMC will allow us to better constrain the nature of dark matter and properties of the LMC.

The plan of the paper is as follows. 
Sec.~\ref{sec:methods} describes our methodology including an overview of BFEs, the Galactic potentials used in this study and the framework to generate stellar streams. 
In Sec.~\ref{sec:results}, we present the distribution of stream statistics for the population of MW streams made in each potential. Further, we demonstrate the detectable signatures of the LMC and MW halo response seen in the stream properties. 
In Sec.~\ref{sec:res-pms}, we compare our results for proper motion misalignments with observational data.
In Sec.~\ref{sec:discussion} we discuss our results in context and assess any caveats. Finally, we summarise and conclude our results in Sec.~\ref{sec:conclusions}.

\section{Methods} \label{sec:methods}

\begin{figure*}
    \centering
    \includegraphics[width=\linewidth]{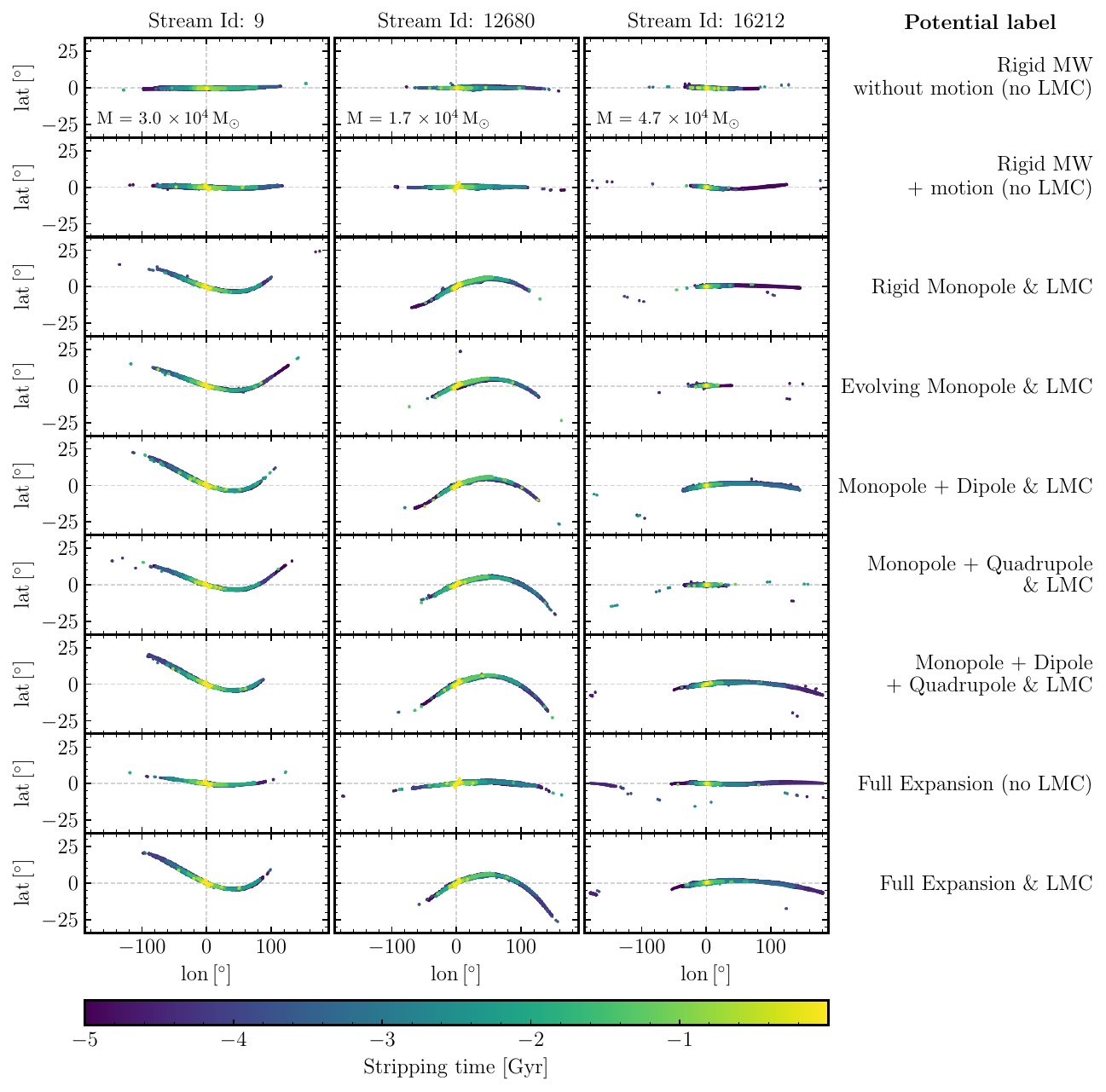}
    \caption{A selection of stellar streams generated using the present-day phase space coordinates drawn from an ergodic distribution function and a progenitor mass assigned from a log-normal mass function, see Sec.~\ref{sec:methods-DFs}. 
    Each column shows an individual stream plotted in the coordinate frame centered on its progenitor, Sec.~\ref{sec:methods-dynmodels}. The stream Id is the column title and the total stellar mass is shown in the top row panels. 
    The x-axis represents the longitude along the stream relative to the progenitor, such that positive/negative longitude corresponds to the leading/trailing tidal tail. 
    The y-axis represents the latitude above the plane of the progenitor's orbit. 
    Each row corresponds to the stream generated in a MW--LMC potential as labelled on the right hand side, Sec.~\ref{sec:methods-galpot}. The colourmap shows the time at which the particles in the stream were tidally stripped from the progenitor relative to the present-day, $t = 0\,\mathrm{Gyr}$. Each column illustrates the diversity in present-day stream morphology given the choice of Galactic potential.}
    \label{fig1}
\end{figure*}

To generate stellar stream models in a time-evolving MW--LMC system, we need to describe the potential, density and forces at any arbitrary position and time. Rigid, non-time evolving, potentials fail to capture deformations to the MW and LMC dark matter haloes. 
We use BFEs, described in Sec.~\ref{sec:methods-bfes}, to expand and evolve the N-body models of the MW and LMC as outlined in Sec.~\ref{sec:methods-Nbody}. Using the BFE description for the Galactic potential, we generate populations of globular cluster streams; Sec.~\ref{sec:methods-gcpops}.

\subsection{Generating a flexible potential model}\label{sec:methods-genpotmodel} 

\subsubsection{Basis Function Expansions}\label{sec:methods-bfes}

BFEs offer a framework to describe these time-dependent deformations. They track the gravitational potential, density, and forces as the system evolves over time. BFEs have previously been seen to accurately describe flexible models of the MW \citep{2016MNRAS.463.1952P, 2018ApJ...858...73D, 2019MNRAS.490.3616P, 2020MNRAS.494L..11P, 2021ApJ...919..109G}. In this work, we use the BFEs of the MW--LMC system presented in \citet{Lilleengen2023} that are simulated using \textsc{exp} \citep{2022MNRAS.510.6201P}, with the expansion coefficients recorded at each time step.

The BFE technique uses appropriately chosen biorthogonal density-potential pairs of basis functions, $\{ \varrho_{\mu}(\textbf{x}), \phi_{\mu}(\textbf{x}) \}$, that solve Poisson's equation, $\nabla^{2}\phi_{\mu}(\textbf{x}) = 4\pi G\varrho_{\mu}(\textbf{x})$, and satisfy the biorthogonality condition, $\int \mathrm{d}^3\mathbf{x}\,\phi_{\mu}(\textbf{x})\varrho_{\nu}(\textbf{x}) = 4\pi G\delta_{\mu \nu}$, where $\delta_{\mu \nu}$ is the Kronecker delta. Each basis function, labeled by the index $\mu$, adds a degree of freedom to the system and has an associated coefficient $A_{\mu}$, which determines its contribution to the total description of the system given by the summation over all basis function terms. A system at any given time is described by the basis functions and the coefficients that weight them. Mathematically, the density, $\rho$, and gravitational potential, $\Phi$, are given by, $\rho(\textbf{x}, t) = \sum_{\mu}\,A_{\mu}(t) \varrho_{\mu}(\textbf{x})$ and $\Phi(\textbf{x}, t) = \sum_{\mu}\,A_{\mu}(t) \phi_{\mu}(\textbf{x})$, where the basis coefficients are time-dependent and the basis functions keep their fixed functional form.

Basis functions are selected to reflect the system they describe. To model density profiles, $\rho(r, \phi, \theta)$, with deviations away from spherical symmetry, the spherical harmonics $Y_{l}^{m}$ are chosen to describe the distribution in the angular coordinates $(\phi, \theta)$, whilst \textsc{exp} describes the radial dependence (index $n$) by the eigenfunctions of a Sturm-Liouville equation \citep{Weinberg1999}. Each basis function is then represented by the triplet of indices $\mu \equiv (n, l, m)$. The radial index, $n$, determines the number of nodes in the radial basis function. 
It is often convenient to describe individual harmonic subsets of $l$. The $l=0$ terms are called the monopole, $l=1$ is the dipole, and $l=2$ is the quadrupole.
The \textsc{exp} method creates a lowest-order monopole term, $\rho_{000}(r)$, that exactly matches the unperturbed, spherical, input potential-density pair. All other, higher-order, terms are perturbations around the input distribution. 
Another example of a BFE is the classical Hernquist-Ostriker basis set \citep{1992ApJ...386..375H} which expands upon the Hernquist density distribution \citep{Hernquist1990} as $\rho_{000}(r)$. This expansion has been used extensively in the literature \citep[e.g.][]{Johnston2001, Johnston2002b, Johnston2002a, Sanders2020}. Alternative choices of analytic basis functions based upon the Navarro-Frenk-White profile \citep[NFW;][]{Navarro1997} have been made such that flattened density distributions are more accurately described \citep{Lilley2018b,Lilley2018a}. 

\subsubsection{\emph{N}-body models}\label{sec:methods-Nbody}

An efficient, lightweight Python interface, \textsc{mwlmc}, has been developed to facilitate the \textsc{exp} simulations of the \citet{Lilleengen2023} MW--LMC system. This user-friendly interface is publicly available at:~\url{https://github.com/sophialilleengen/mwlmc}. This MW--LMC system is made up of three constituents with separate BFEs: the MW dark matter halo, the MW stellar disc and bulge and the LMC dark matter halo. The \textsc{exp} method explicitly uses the BFE for the force evaluations in the $N$-body evolution. We summarise all components in Table.~\ref{table1}. Detailed descriptions of the BFE and $N$-body models can be found in secs.~2.1 \& 2.2 of \citet[][]{Lilleengen2023}, respectively.

The live simulation of the MW--LMC system begins at $t = -2.5\,\mathrm{Gyr}$, with present-day at $t = 0\,\mathrm{Gyr}$. The simulation aims to set the 6D present-day LMC coordinates with:~a right ascension and declination of ($\alpha$, $\delta$) = ($78.76^{\circ} \pm 0.52,\, - 69.19^{\circ} \pm 0.25$), 
proper motions of $(\mu_{\alpha}$, $\mu_{\delta})$ = ($1.91 \pm 0.02, 0.229 \pm 0.047$) $\,\mathrm{mas}\,\mathrm{yr}^{-1}$ \citep{Kallivayalil2013}, 
a distance of $d = 49.59 \pm 0.54 \,\mathrm{kpc}$ \citep{Pietrzynski2019},
and a line of sight velocity of $v_{\mathrm{los},} = 262.2 \pm 3.4\,\mathrm{km}\,\mathrm{s}^{-1}$ \citep{vanderMarel2002}. 
At the start of the live simulation, the MW and LMC haloes are totally distinct, with the LMC outside the virial radius of the MW at a distance of $450\,\mathrm{kpc}$.
The density, force, and potential fields before the start of the live simulation have the basis coefficients set to their initial values prescribed at $t_{\mathrm{live}}$.

\subsection{Globular cluster populations in MW--LMC systems}\label{sec:methods-gcpops}

\setlength{\tabcolsep}{3pt}
\begin{table*}
\centering
\caption{Summary table for the Galactic potentials used throughout this study, Sec.~\ref{sec:methods-galpot}. The colour index corresponds to the line colours used to represent each potential in figures.}
\begin{tabular}{ccccccccc} 
\hline
\hline
Potential label & \shortstack{\\ Colour \\ index} & MW halo & $l_{\mathrm{exp,\,halo}}$ & MW disc & $m_{\mathrm{exp,\,disc}}$ & \shortstack{\\ Reflex \\ motion} & LMC & $l_{\mathrm{exp,\,LMC}}$\\
\hline
\hline
 \textit{Rigid MW without motion (no LMC)} & \textcolor{pot1}{\rule{.5cm}{1.5mm}} &\ding{51} (Rigid) & 0 & \ding{51} (Rigid) & 0 & \ding{55} & \ding{55} & - \\
\textit{Rigid MW + motion (no LMC)} & \textcolor{pot2}{\rule{.5cm}{1.5mm}} & \ding{51} (Rigid) & 0  & \ding{51} (Rigid) & 0 & \ding{51} & \ding{55} & -\\
\textit{Rigid Monopole \& LMC} & \textcolor{pot3}{\rule{.5cm}{1.5mm}} & \ding{51} (Rigid) & 0  & \ding{51} (Live) & 0-6 &\ding{51} & \ding{51} (Live) & 0-6\\
\textit{Evolving Monopole \& LMC} & \textcolor{pot4}{\rule{.5cm}{1.5mm}} & \ding{51} (Live) & 0  & \ding{51} (Live) & 0-6 & \ding{51} & \ding{51} (Live) & 0-6\\
\textit{Monopole + Dipole \& LMC} & \textcolor{pot5}{\rule{.5cm}{1.5mm}} & \ding{51} (Live) & 0, 1  & \ding{51} (Live) & 0-6 & \ding{51} & \ding{51} (Live) & 0-6\\
\textit{Monopole + Quadrupole \& LMC} & \textcolor{pot6}{\rule{.5cm}{1.5mm}} & \ding{51} (Live) & 0, 2  & \ding{51} (Live) & 0-6 & \ding{51} & \ding{51} (Live) & 0-6\\
\textit{Monopole + Dipole + Quadrupole \& LMC} & \textcolor{pot7}{\rule{.5cm}{1.5mm}} & \ding{51} (Live) & 0, 1, 2  & \ding{51} (Live) & 0-6 & \ding{51} & \ding{51} (Live) & 0-6\\
\textit{Full Expansion \& LMC} & \textcolor{black}{\rule{.5cm}{1.5mm}} & \ding{51} (Live) & 0 - 6  & \ding{51} (Live) & 0-6 & \ding{51} & \ding{51} (Live) & 0-6\\
\textit{Full Expansion (no LMC)} & \textcolor{black}{\rule{.2cm}{1.5mm}} \textcolor{black}{\rule{.2cm}{1.5mm}} & \ding{51} (Live) & 0 - 6  & \ding{51} (Live) & 0-6 & \ding{51} & \ding{55} & -\\
\hline
 \label{table2}
\end{tabular}
\end{table*}

\subsubsection{Initial conditions - Present-day globular cluster distribution}\label{sec:methods-DFs}

To generate the present-day phase space coordinates of our globular cluster sample, we draw samples from an ergodic distribution function implemented in \textsc{agama} \citep{2019MNRAS.482.1525V}. This requires instances of a tracer density and potential profile. We use a Dehnen tracer density profile  \citep{Dehnen1993} and an NFW profile for the potential \citep{Navarro1997} that matches the rigid monopole order expansion of the MW halo potential used in \citet{Lilleengen2023}; see also our Sec.~\ref{sec:methods-Nbody}. We over-sample this distribution function to return a set of $3\times 10^5$ possible present-day phase space coordinates.

We impose certain criteria on the allowed range of orbits given the returned phase-space positions and velocities. We enforce the orbits to have a pericenter between $10\,\mathrm{kpc}$ and $25\,\mathrm{kpc}$, and an apocenter no greater than $75\,\mathrm{kpc}$. The cut on the allowed orbital pericenter is to produce streams with sufficiently disrupted and extended tidal tails. Streams with pericenters smaller than this range begin to wrap and span the full longitudinal axis in progenitor coordinates, see Sec.~\ref{sec:methods-dynmodels}. This makes our simple stream statistics in Sec.~\ref{sec:sumstats} difficult to interpret. Further, allowing streams to orbit within heliocentric distances requires careful distinction between galacto-centric and helio-centric quantities. The cut on the allowed progenitor apocenter, and pericenter, will limit the analysis to interesting streams that interact with the LMC near its pericenter and so do not spend large portions of time at very large Galactocentric radii.
To determine the peri-/apo-centers for each final condition drawn from the distribution function, we integrate their orbits for $5\,\mathrm{Gyr}$ using \textsc{gala} \citep{gala}. For the Galactic potential, we use the same prescription as used to instantiate the distribution function. These calculated peri-/apo-centers may differ from those calculated using a time-evolving potential. Of the orbits which pass these criteria, we take a random sample of $16,384$ phase-space positions and velocities to be our present-day distribution of stream progenitors.  

We model the density profile for each globular cluster as a Plummer sphere \citep{1911MNRAS..71..460P}. This requires knowledge of the mass and scale radius of the cluster. To assign a mass to each cluster, $M_{\mathrm{prog}}$, we randomly draw values from the initial log-normal mass function shown in \citet{Gnedin2014}, fig.~2. The scale radius, $a_s$, is set to be $2\,\mathrm{pc}$ for every globular cluster \citep[a suitable choice for globular clusters e.g., the GD-1 globular cluster stream,][]{Dillamore2022}. We choose to consider only globular cluster streams as, in the absence of external perturbations, they form dynamically cold and angularly (on-sky) thin stellar streams. These streams highlight stream perturbations due to the LMC and/or MW halo response more clearly than dynamically hotter dwarf galaxy streams.

\subsubsection{Dynamical modelling of stellar streams}\label{sec:methods-dynmodels} 

We model stellar streams using a ‘modified Lagrange Cloud Stripping' (mLCS) technique \citep{Kupper2012, Bonaca2014, 2014MNRAS.445.3788G, 2015MNRAS.449.1391B, 2015MNRAS.452..301F}. Modifications were developed to include the forces from the LMC and the reflex motion of the MW in \citet{2019MNRAS.487.2685E}. The stream progenitors are modelled as Plummer spheres \citep{1911MNRAS..71..460P} with initial masses and scale radii as defined in Sec.~\ref{sec:methods-DFs}. Our mock stream implementation is similar to that in \citet{Brooks2024}. We make sure to integrate orbits in the inertial frame of reference. By integrating orbits in non-inertial frames of reference, this can result in fictitious, i.e. non-physical, forces acting on streams \citep[see,][appendix A for discussion on inertial forces in cosmological simulations]{Arora2024}. This can lead to biases in stream fitting used to evaluate the Galactic potential. We discuss frames of reference and fictitious forces in Appendix~\ref{app:A}.

From the progenitor's present-day position, we rewind the phase-space orbit for $5\,\mathrm{Gyr}$ in a chosen MW--LMC potential; see Sec.~\ref{sec:methods-galpot}. The system is subsequently forward-evolved in the same potential, and stream particles are released from the progenitor's Lagrange points, $r_{\mathrm{prog}} \pm r_t$, at each time step. 
The Lagrange, or tidal, radius is found by, $r_{t} = \left( \mathrm{G}M_{\mathrm{prog}}(t) / ( \omega^{2} - \mathrm{d}^2\Phi/\mathrm{d}r^2) \right)^{1/3}$, where $\omega$ is the angular velocity of the progenitor with respect to the MW and $\mathrm{d}^2\Phi/\mathrm{d}r^2$ is the second derivative of the MW potential along the radial direction. Tidal stripping from the LMC is ignored.
We model the mass loss of progenitors as linearly decreasing in time, such that they are fully disrupted at present-day, since many progenitors are not seen in observational data at present-day \citep{Koposov2023}.
We account for the velocity dispersion of the progenitor, $\sigma_{\mathrm{v}}$, by randomly drawing velocities from a 3D isotropic Gaussian centred on the velocities of the stripped particles, $\textbf{v}_{\mathrm{strip}}$, with standard deviation $\sigma_{\mathrm{v}} = \sqrt{\mathrm{G}M_{\mathrm{prog}}(t)/(r_t^2 + a_s^2)^{1/2}}$, where $a_s$ is the scale radius of the progenitor. The radial component of $\textbf{v}_{\mathrm{strip}}$ is the same as the progenitor, while the tangential components are set equal to those at the point
halfway between the position of the progenitor and the Lagrange point.

To visualise the resulting stream, we rotate it into its own stream coordinates such that its progenitor lies at $(0^{\circ}, 0^{\circ})$. To do this, we rotate around the $z$ and $y$ axes by the progenitor's longitude and latitude in Galactic coordinates, respectively. This defines a new $x$-axis about which we do a final rotation by the angular displacement of the progenitor's velocity vector. In Fig.~\ref{fig1} we show this for three example streams generated in each potential we consider, Sec.~\ref{sec:methods-galpot}.

\subsubsection{Galactic potentials}\label{sec:methods-galpot} 

We are interested in determining the effect on streams of direct LMC interactions and by the individual contributions to the full BFE description of the MW halo response. For the latter, we can selectively turn off harmonic terms of the full BFE, i.e. by setting all relevant BFE coefficients to zero, to isolate a given contribution to the total BFE description of the system. 
In this work, we consider 9 unique potentials in which we generate a population of stellar streams. Table~\ref{table2} details the Galactic potentials used in this study. This table describes which individual potential components are included in the total potential, Sec.~\ref{sec:methods-Nbody}. Additionally, we state whether each component is rigid or allowed to deform, and if so, which harmonic terms are included in the basis expansion. We also provide labels for each potential we use throughout this study and the associated colour used in the figures.

\begin{figure*}
    \centering
    \input{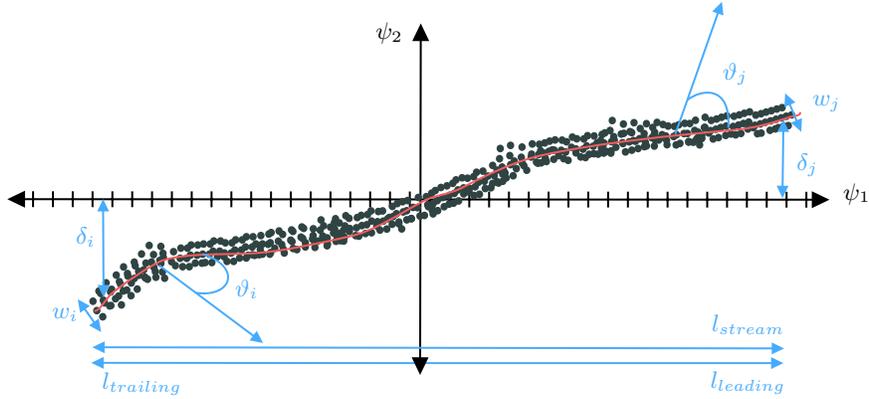} 
    \caption{Illustration of a how we measure the widths, $w$, lengths, $l_{\mathrm{stream}}$, stream track deviation, $\delta$ and proper motion misalignment angle, $\vartheta$ as outlined in Sec.~\ref{sec:sumstats}. Subscript notation indicates we measure each quantity in the $i^{\mathrm{th}}$ to the $j^{\mathrm{th}}$ bin in $\psi_1$. 
    The axis represents the longitude, $\psi_1$ and latitude $\psi_2$ of stream particles relative to the progenitor at $(0,0)$. The Great Circle track is equivalent to $\psi_1 = 0^{\circ}$. Example stream particles are shown as grey dots with the red line through them representing the stream track. }
    \label{fig2}
\end{figure*}

\section{Results} \label{sec:results}

\subsection{Intuition and expectations}\label{sec:res-expectations} 

The orbits of stellar streams in spherical potentials can be described entirely by their energy and angular momentum \citep{Johnston1998, Helmi2020}. For a Galactocentric frame of reference, the orbit of the stream will be contained to a single plane as angular momentum is conserved \citep{Mateu2023}; this is the Great Circle orbit. Equivalently, in orbital pole space, as all stream stars will share similar and constant angular momentum, the spread in poles will be minimal. 
In the absence of perturbations, stream stars roughly delineate orbits that follow the stream progenitor \citep{McGlynn1990, Johnston1996, Binney2008, Sanders2013}. This implies their velocities, or proper motions in projection, are aligned with the stream track.
Furthermore, the length of a stream is set by the time for which the progenitor has been tidally disrupting and its orbital frequency. The orbital frequencies vary along a stream such that the stars in the leading stream tail will have greater orbital frequencies than those in the trailing tail. Over time, this will cause a stream to be slightly asymmetric with bias towards longer leading tidal tails. 
The width of a stream is set by the scale over which debris is distributed; the tidal radius \citep{Johnston2016}. Moreover, density variations along a stream reflect the number and frequency of visits to pericenter/apocenter. At pericenter, tidal forces are maximal and the most debris will be stripped from the progenitor. Conversely, the opposite is true at apocenter. This leads to periodic, or ‘epicyclic', over- and under-densities of stars along a stream equating to pericenter and apocenter passages \citep{Kupper2008, Kupper2010, Kupper2012}. By generating streams using a linear time-step for orbit integration, see Sec.~\ref{sec:methods-dynmodels}, we do not produce epicyclic density behaviour. Nevertheless, this modelling choice does not alter any of our analysis.

When the Galactic potential becomes more complex, e.g., for time-independent axisymmetric, or triaxial, MW dark matter halo shapes, this leads to more complicated stream morphologies and evolution. In non-spherical potentials, a stream will broaden due to precession of the orbital planes of stream stars as the total angular momentum is no longer conserved \citep{Erkal2016a}. Indeed, if the Hessian of the potential in action-angle coordinates has more than one eigenvalue with comparable magnitudes, a stream will form a multi-dimensional structure as opposed to the simple one dimensional structure \citep{Sanders2013}. For example, streams close to the Galactic disc exhibit a range of vertical orbital frequencies forming two-dimensional ‘ribbon' like structures \citep{Dehnen2018}. Additionally, for streams orbiting in regions of resonance, this further complicates the morphology of streams e.g., ‘fans' of tidal debris for the Pal-5 stream \citep{Pearson2015, Price-Whelan2016, Yavetz2021, Yavetz2023}.

Interactions of streams with time-dependent structures in the Galaxy, such as the MW bar, can cause more rapid diffusion of streams \citep{Price-Whelan2016}, truncate the length of streams \citep{Hattori2016} and lead to fanning of tidal debris \citep{Pearson2017}. The interactions of streams with dark matter subhaloes can cause ‘gaps' in streams to develop due to impulsive velocity kicks imparted on the stream \citep{Ibata2002, Johnston2002a, Yoon2011, Carlberg2012, Erkal2015a, Sanders2016}. Moreover, the interactions of streams with massive dwarf galaxies can affect their morphology e.g. causing stream length asymmetries in the GD-1 stream \citep{Dillamore2022}. As the most recent dwarf galaxy merger, the LMC has caused disequilibrium in the MW \citep[e.g.,][]{Belokurov2019, Garavito-Camargo2019, 2019MNRAS.487.2685E, 2020MNRAS.494L..11P, Vasiliev2021a, Conroy2021, Petersen2021, Erkal2021, Amarante2024, Yaaqib2024, Chandra2024}. The effect of the LMC on individual streams has been observed e.g., Orphan-Chenab \citep{Koposov2019, Erkal2019, Shipp2019, Shipp2021}, but the general effect on the entire population of MW streams remains to be answered. In the following sections, we address how the morphology and other stream observables are affected by their interactions with the LMC and/or via the MW halo response.

\subsection{Summary statistics}\label{sec:sumstats} 

To quantify how the choice of Galactic potential affects the resulting stream population, we use a set of Galactocentric summary statistics motivated by the expectations above, Sec.~\ref{sec:res-expectations}. We categorise these statistics into three parts: properties of the stream initial conditions, dynamical analysis and direct observables. The initial condition statistics are defined by:

\begin{description}
    \item[Progenitor mass] $M_{\mathrm{prog}}\,[\mathrm{M}_{\odot}]$, the mass of the progenitor before any tidal stripping occurs. This mass is also the total mass of the stellar stream at present-day.
    \item[Pericenter] $r_p\,[\mathrm{kpc}]$, the radial distance from the center of the MW when the stream is at its closest position along its orbit. Calculated when integrating orbits in the time-independent potential, Sec.~\ref{sec:methods-DFs}.
    \item[Apocenter] $r_a\,[\mathrm{kpc}]$, the radial distance from the center of the MW when the stream is at its furthest position along its orbit. Calculated when integrating orbits in the time-independent potential, Sec.~\ref{sec:methods-DFs}.
\end{description}

\noindent
For the dynamical analysis, we use the following statistics defined as:

\begin{description}
    \item[Energy] $E \, [(\mathrm{km}\,\mathrm{s}^{-1})^2]$, the Galactocentric energy of stream members. The energy for each stream particle is defined to be the sum of the kinetic plus potential energies.
    \item[Angular momenta] $(L,\,L_z) \, [\mathrm{km}\,\mathrm{s}^{-1}\,\mathrm{kpc}]$, the Galactocentric total and z-direction angular momenta of the stream particles. 
    \item[Orbital poles] $(l^{\prime}_{\mathrm{pole}}\,,\,b^{\prime}_{\mathrm{pole}}) \, [^{\circ}]$, using the stream particle positions and velocities in the Galactocentric Cartesian frame, we find the angular momentum vector as the cross product of the position and velocity vectors. We decompose the resulting vector into a normalised direction; the orbital pole, i.e $\mathbf{J} = \mathbf{r} \times \mathbf{v} /\,|\,\mathbf{r} \times \mathbf{v}|$. We rotate the orbital poles with respect to the orbital pole direction of the MW disc which we arbitrarily place at $(l^{\prime}, b^{\prime}) = (0^{\circ}, 90^{\circ})$. 
\end{description}

\noindent
And, for the stream observables, use the following statistics, illustrated in Figure \ref{fig2}, defined as:
\begin{description}
    \item[(Angular) length] $l_{\mathrm{stream}}\,[^{\circ}]$, the length of the stream in degrees. Taken between the $5^{\mathrm{th}}$ and $95^{\mathrm{th}}$ percentile along the longitude in stream coordinates. 
    \item[Asymmetry] $l_{\mathrm{leading}}/l_{\mathrm{trailing}}$, the ratio of the length of the stream leading and trailing tidal arms measured in degrees. For a perfectly symmetric stream this ratio is equal to 1. A ratio above/below this value indicates a stream that is preferentially longer in its leading/trailing tidal arms.
    \item[(Angular) width] $w\,[^{\circ}]$, the width of the stream in degrees. We take 50 bins in the longitudinal stream coordinates, $\psi_1$. For each bin, we find the standard deviation of the corresponding latitudes, $\psi_2$, for stream particles. The width is then taken as the median value across all bins.
    \item[Deviation from Great Circle] $\bar{\delta}\,[^{\circ}]$, the deviation of the stream in degrees from the unperturbed stream track i.e. a Great Circle orbit. We take 50 bins in the longitudinal stream coordinates. For each bin, we find the absolute median latitude in stream coordinates; see Fig.~\ref{fig2}. As $\psi_2 = 0\,^{\circ}$ corresponds to the Great Circle orbit, this angular value represents any deviation due to time-dependence in the potential. We quote the median value across all bins. 
    \item[Local velocity dispersion]  $\sigma_v \, [\mathrm{km}\,\mathrm{s}^{-1}]$, the velocity dispersion of the stream across the width of the stream. We take 50 bins in the longitudinal stream coordinates between the $5^{\mathrm{th}}$ and $95^{\mathrm{th}}$ percentiles. For each bin, we find the standard deviation of stream particle velocities. The local velocity dispersion is taken as the median value across all bins.
    \item[Proper motion misalignment] $\bar{\vartheta} \, [^{\circ}]$, following \citet{Mateu2023} we measure the angular separation between the angular momentum and the pole vector along the track. The misalignment angle is reported as the median value across the stream. The pole track is computed as the vector product between the adjacent position vectors along a stream. This pole track may change along the track if the stream is not perfectly planar. This is equivalent to \citet{Erkal2019} who compare the tangent to the stream track with the ratio of proper motions along the track.
    \item[Orbital pole dispersion] $(\sigma_{l,\,\mathrm{pole}}, \,\sigma_{b,\,\mathrm{pole}}) \, [^{\circ}]$, the dispersion is calculated as the standard deviation of the orbital poles calculated for each stream particle as described in the preceding bullet.
\end{description}

\begin{figure*}
    \centering
    \includegraphics[width=\linewidth]{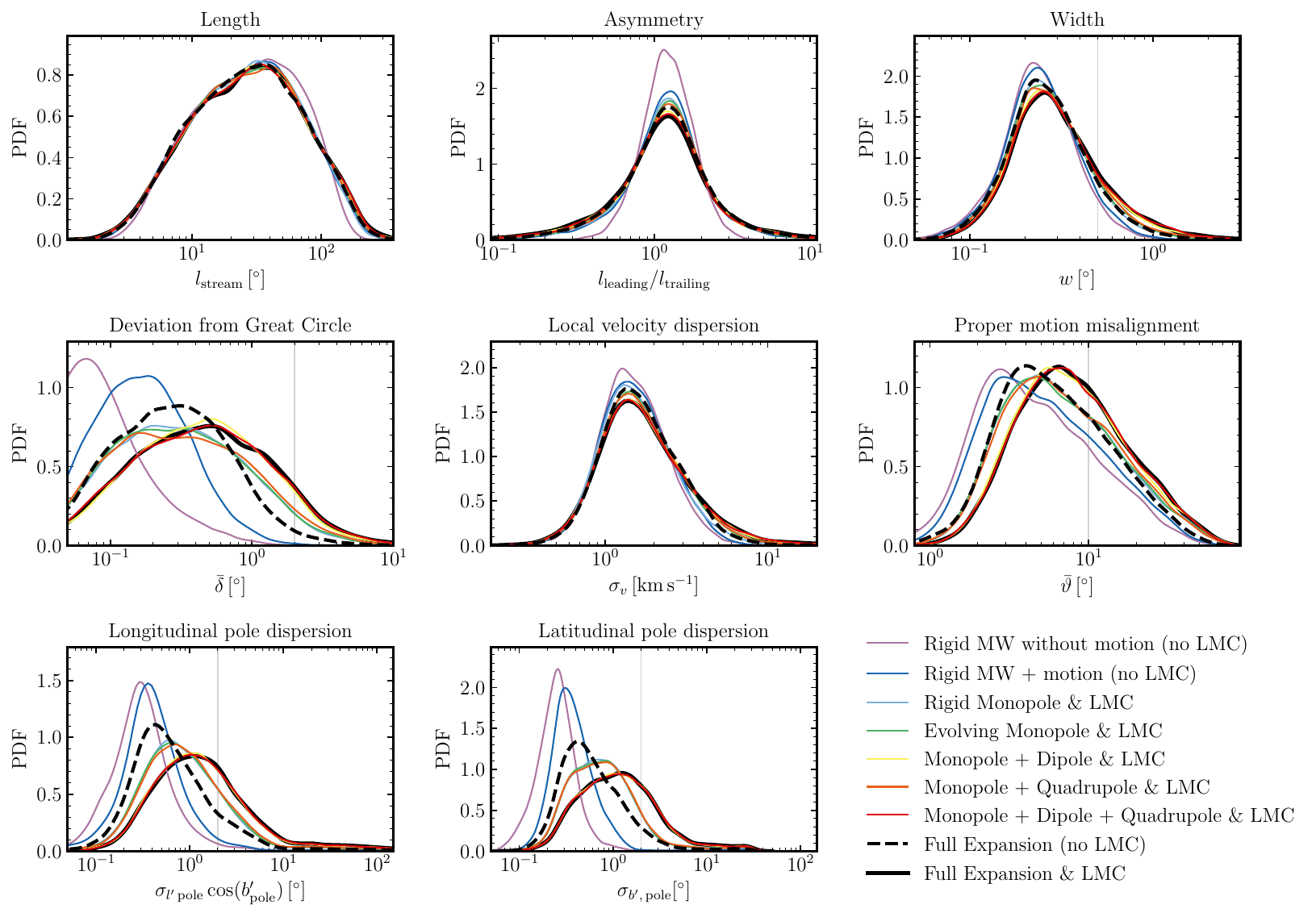}
    \caption{Probability density functions, inverse units of x-axes, for a selection of statistics that summarise stream observables, Sec.~\ref{sec:sumstats}. Left to right, and top to bottom, and we show the length, $l_{\mathrm{stream}}$, 
    the asymmetry of the stream tidal tails, $l_{\mathrm{leading}}/l_{\mathrm{trailing}}$,
    the width, $w$, 
    the track deviation from the progenitor's Great Circle orbit, $\bar{\delta}$,
    the local velocity dispersion, $\sigma_{v}$,
    the proper motion misalignment angle $\bar{\vartheta}$, 
    and the spreads in the longitudinal and latitudinal orbital poles rotated into the MW disc's orbital pole frame, $\sigma_{l^{\prime},\,\mathrm{pole}}$ and $\sigma_{b^{\prime},\,\mathrm{pole}}$. For each statistic, we plot lines associated with each potential where we have generated a population of streams in, see Table.~\ref{table2} for details. The thick solid (dashed) black line shows the MW--LMC potential that is described by a full basis expansion MW with (without) a full basis expansion LMC. The statistics with the most striking differences between the potential used to generate the streams are the widths, track deviations, proper motion misalignments, and orbital pole spreads. For these statistics, we draw a vertical grey line corresponding to the ‘detectable' threshold value we use to investigate LMC or MW halo response effects in Sec.~\ref{sec:res-obs-effects}.}
    \label{fig3}
\end{figure*}

\begin{figure*}
    \centering
    \includegraphics[width=0.8\linewidth]{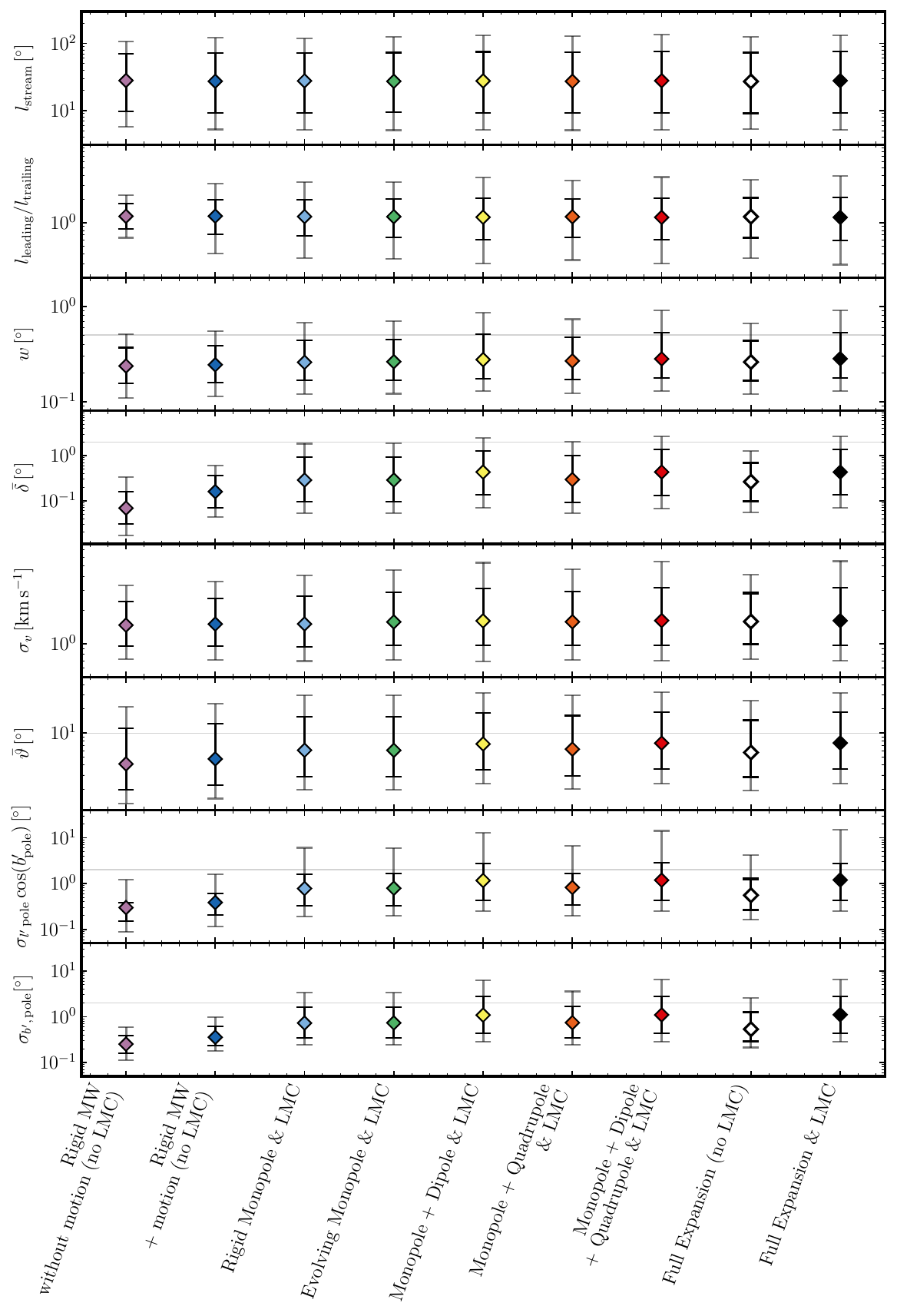}
    \caption{The median, $16^{\mathrm{th}}-84^{\mathrm{th}}$ (black error bars) and  $5^{\mathrm{th}}-95^{\mathrm{th}}$ (grey error bars) percentiles of the stream statistic probability density functions in Fig.~\ref{fig3} for each potential used to generate a population of streams; see Table~\ref{table2} for colours and description. Top to bottom we show the length, $l_{\mathrm{stream}}$, 
    the asymmetry of the stream tidal tails, $l_{\mathrm{leading}}/l_{\mathrm{trailing}}$,
    the width, $w$, 
    the track deviation from the progenitor's Great Circle orbit, $\bar{\delta}$,
    the local velocity dispersion, $\sigma_{v}$,
    the proper motion misalignment angle $\bar{\vartheta}$, 
    and the spreads in the longitudinal and latitudinal orbital poles rotated into the MW disc's orbital pole frame, $\sigma_{l^{\prime},\,\mathrm{pole}}$ and $\sigma_{b^{\prime},\,\mathrm{pole}}$. We plot horizontal grey lines corresponding to the ‘detectable' threshold value we use to investigate LMC or MW halo response effects in Sec.~\ref{sec:res-obs-effects}.}
    \label{fig4}
\end{figure*}

\subsubsection{Distributions of summary statistics}\label{sec:sumstats-dists} 

\begin{figure*}
    \centering
    \includegraphics[width=\linewidth]{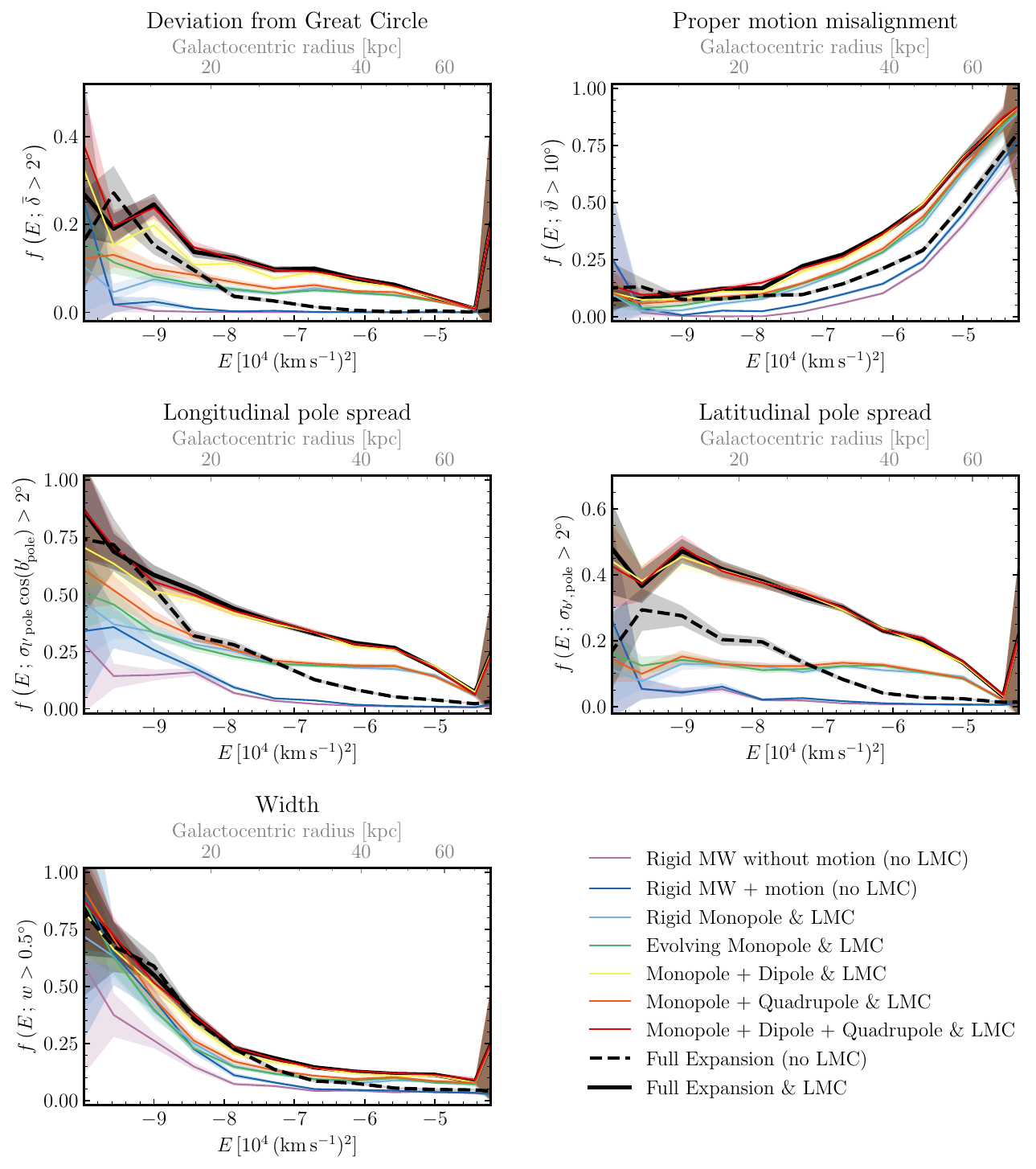}
    \caption{Fraction of streams with ‘\textit{detectable}' signatures of the LMC as a function of the stream progenitor's energy (lower x-axis), or equivalently, radial Galactocentric position (upper x-axis). For each panel, a ‘detectable' signature in a stream at present-day constitutes one in which: 
    \textit{top left:} the deviation from the Great Circle track exceeds $2^{\circ}$; 
    \textit{top right:} the proper motion misalignment exceeds $10^{\circ}$;
    \textit{middle left:} the spread in the longitudinal pole rotated into the MW disc's orbital pole frame exceeds $1^{\circ}$;
    \textit{middle right:} the spread in the latitudinal pole rotated into the MW disc's orbital pole frame does not exceed $1^{\circ}$; and 
    \textit{lower left:} the stream width exceeds $0.5^{\circ}$. We plot lines associated with each potential that we generate a population of streams in, see Table.~\ref{table2}. The thick solid (dashed) black line shows the MW--LMC potential that is described by a full basis expansion MW with (without) a full basis expansion LMC. Poisson error bars are shown for each potential as the shaded region of the same colour. In the inner/outer halo, the MW halo response / LMC dominates the effect on streams.}
    \label{fig5}
\end{figure*}

To analyse how the choice of Galactic potential affects the properties of a population of stellar streams, we plot the probability density functions in Fig.~\ref{fig3} for the statistics that describe stream observables, Sec.~\ref{sec:sumstats}. In Fig.~\ref{fig4} we plot the median, $16^{\mathrm{th}}-84^{\mathrm{th}}$ (black error bars) and  $5^{\mathrm{th}}-95^{\mathrm{th}}$ (grey error bars) percentiles of these probability density functions. 
In Fig.~\ref{fig3}, we plot the fiducial ‘Full Expansion \& LMC' potential as the thick solid black line and the ‘Full Expansion (no LMC)' as the thick dashed black line. By comparing these two potentials, we physically separate the direct influence of the LMC's gravity from the indirect effects of the MW halo response. The coloured lines represent the distributions associated with potentials that use a subset, or rigid versions, of the full expansion MW BFEs; see Table.~\ref{table2} for details. Relative comparison of these potentials will isolate effects resulting from harmonics of the MW halo response. For example, the comparison of the ‘Evolving Monopole \& LMC' and ‘Monopole + Dipole \& LMC' potentials isolates the importance of the MW dipole. Physically the dipole is attributed to the displacement of the MW center of mass.

We find that some stream statistics are largely independent of the choice of Galactic potential used to generate the stream population. For example, the stream length, asymmetry, and velocity dispersion display similar distributions regardless of the potential. In Fig.~\ref{fig4}, this can be seen by the median and/or percentile boundaries being largely unaffected between potentials. The exception to this can be seen when the reflex motion of the MW barycenter is included. This effect lengthens some streams, increases the number of more significantly asymmetric streams, and increases the stream velocity dispersion. 

We find that the stream statistics with clearer dependencies on the choice of Galactic potential used to generate a population of streams are:~the widths, track deviations, proper motion misalignments, and orbital pole spreads. In Fig.~\ref{fig4}, this is seen by the median and/or percentile boundaries changing between the potentials.
By comparing the ‘Full Expansion \& LMC' and the ‘Full Expansion (no LMC)' potentials, we can see the direct perturbation of the LMC as increasing the median value of these statistics.
Moreover, the greatest change by response of the MW halo to the distributions is via its dipole harmonic.
For these statistics, we want to find which are the best indicators for perturbations due to the LMC and the MW halo response, both as a function of Galactocentric distance and the on-sky angular position. 

\subsubsection{Detectable stream perturbations}\label{sec:res-detectable-signatures}

For the stream statistics dependent on the Galactic potential:~widths, track deviations, proper motion misalignments, and orbital pole spreads, we propose ‘detectable' threshold values to capture the perturbations due to the LMC and the MW halo response. These threshold values are motivated by the expected observational values a stellar stream would display if there had been no external perturbations, i.e. stream properties are solely due to the intrinsic, physical processes that generate the stream itself. Hence, for streams that have summary statistic values greater than these threshold values, we associate these streams as being perturbed either directly by the LMC or by the response of the MW halo.

For a stream's width, this is controlled by the tidal radius which depends on the Galactocentric distance and the masses of the progenitor and host galaxy \citep{Johnston1998}. For a globular cluster, $\sim10^{4}\,\mathrm{M}_{\odot}$ in the MW, $\sim10^{12}\,\mathrm{M}_{\odot}$ over the distances we consider, $\lesssim75\,\mathrm{kpc}$, we can expect angular stream widths of $\sim0.5^{\circ}$. 
For the track deviations, we are motivated by the \textit{galstreams} catalog of MW stellar streams \citep{Mateu2023}. The great circle approximation is a good one when studying outer halo streams, $\gtrsim25\,\mathrm{kpc}$, in a Galactocentric frame of reference. The \textit{galstreams} catalog shows that more distant streams generally have smaller off-stream-plane deviations; fig.~15 of \citet{Mateu2023}. Assuming the majority of the known most distant streams are unperturbed, we adopt a threshold value of $2^{\circ}$ to indicate perturbation due to an external source.
For the proper motion misalignments, we are motivated by the results of \citet{Erkal2019} and \citet{Mateu2023}. The former highlights proper motion perturbations to the Orphan-Chenab stream due to LMC. Where the stream is most affected by the LMC, the proper motion misalignment angle is always greater than $\sim10^{\circ}$.
The orbital pole spreads depend on the spread of the positions, from the tidal radius, and velocities, from the velocity dispersion, of a stream. Using the values for these as motivated above, we can expect a pole spread of $\sim1^{\circ}$. For these statistics, we plot a vertical grey line in Fig.~\ref{fig3} corresponding to their ‘detectable' threshold value.

Although, these threshold values are physically and/or observationally motivated, there is still some subjectivity in their choice. The choice of threshold value will impact the amplitude of the fractional number of streams that display ‘detectable' perturbations. 
The most insightful results will come from the relative comparisons between potentials, see Sec.~\ref{sec:res-obs-effects-radialdep}, and as a function of angular position on-sky, see Sec.~\ref{sec:res-obs-effects-angdep}. For small changes in the choice of threshold value, the qualitative result of this study remain unchanged. Furthermore, the Galactic potentials used in this work do not capture all sources of possible perturbation in the MW, see Sec.~\ref{sec:res-expectations}. Although it is beyond the scope of this work, it will be interesting to determine the extent to which other sources of perturbations \citep[e.g., resonances with the MW bar,][]{Dillamore2023, Dillamore2024} affect streams.

\subsection{Trends of stream observables with location}\label{sec:res-obs-effects} 

Sec.~\ref{sec:sumstats-dists} demonstrated clear distinctions between the stream statistics distributions depending on the choice of Galactic potential used to generate the stream population. We want to know whether these distinctions occur in a particular, predictable locations of the Galaxy and whether the LMC and MW halo response effects are discernible.
To achieve this, we adopt the detectable threshold values, as motivated in Sec.~\ref{sec:res-detectable-signatures}, to find the fraction of streams that have been significantly perturbed due to the LMC or the MW halo response. 
In Sec.~\ref{sec:res-obs-effects-radialdep}, we then investigate how the fraction of perturbed streams varies radially outwards through the MW halo. 
Furthermore, in Sec.~\ref{sec:res-obs-effects-angdep}, we investigate the angular dependence of any perturbations to determine whether a local signature of the LMC could be observed.

\subsubsection{Radial dependence}\label{sec:res-obs-effects-radialdep} 

In Fig.~\ref{fig5} we show the fraction of streams with detectable signatures of the LMC or MW halo response effect as a function of the stream progenitor's energy (lower x-axis), or equivalently, the progenitor's radial Galactocentric position (upper x-axis). We interpolate between energy and Galactocentric radius using the circular velocity profile of the MW halo. We plot the fiducial ‘Full Expansion \& LMC' potential as the thick solid black line and the ‘Full Expansion (no LMC)' as the thick dashed black line. The coloured lines represent the distributions associated with potentials that use a subset, or rigid versions, of the full expansion MW and LMC BFEs; see Table.~\ref{table2} for details. The shaded regions around each line represent the Poisson uncertainty.

By comparing the ‘Full Expansion \& LMC' and ‘Full Expansion (no LMC)' potentials, we are able to discern the direct gravitational influence of the LMC on streams. Strikingly, we find for the statistics considered, that there is no overlap of these potential within 1 standard deviation across large parts of the MW. The inclusion of the LMC in the potential significantly increases the fraction of streams with detectable perturbative features by $\sim15 - 25\%$ for the proper motion misalignments ($\sim25 - 70\,\mathrm{kpc}$), and orbital pole spreads ($\sim20 - 50\,\mathrm{kpc}$) stream statistics. Whereas, only a $\sim5\%$ increase is seen in the deviation from a Great Circle orbit ($\sim15 - 45\,\mathrm{kpc}$) and the widths of streams ($\sim25 - 70\,\mathrm{kpc}$). These results imply that for the radial ranges stated, when we average stream statistics across the whole sky, we should be able to directly infer the gravitational impact of the LMC on MW streams given sufficient observational data.

The MW halo response captures how the dark matter halo reacts to the infalling LMC. By isolating individual contributions to the total response of the MW halo, e.g. the dipole harmonic, we are able to investigate how the inclusion of a harmonic mode in the global response affects the perturbations to MW streams. To determine this, we can compare the ‘Rigid Monopole \& LMC' (light-blue line) potential to others that include an LMC but allow for time-variation and/or higher-order harmonics to be included in the MW halo BFE. 
Interestingly, there is no additional contribution to the fraction of streams affected when the monopole harmonic (green line), and/or the quadrupole harmonic (orange line), are allowed to be time-dependent and included in the MW halo expansion. 
Hence, the response of the MW is completely dominated by the inclusion of the dipole harmonic (yellow line; the MW halo dipole captures the reflex and center of mass motions). This increases the fraction of streams affected by up to e.g.~$\sim30\%$ for the latitudinal orbital pole spreads. 
This suggests that the monopole and quadrupole contributions to the MW halo potential are significantly weaker than the dipole, consistent with the findings of \citet{Lilleengen2023} \& \citet{Brooks2024}. Furthermore, any differences due to the MW halo harmonics appear to be systematic over the full Galactocentric distance shown.

The distributions of the various stream statistics in Fig.~\ref{fig5} show contrasting radial profiles. 
For streams residing in the outskirts of the MW $\gtrsim 30\,\mathrm{kpc}$, the LMC has the greatest impact on their proper motions. Indeed, there is a positive correlation between the fraction of streams with detectable proper motion changes and Galactocentric radius. At these distances for the fiducial MW--LMC potential, the fraction of streams with detectable misalignments of their proper motion vectors relative to their stream track increases beyond $50\,\%$. Future surveys like LSST \citep{Ivezic2019} at the \textit{Rubin Observatory} are set to detect many more outer halo streams. As the amount of data increases it will be interesting to see whether these predictions match the observations; see Sec.~\ref{sec:res-pms-lsst}.
For streams in the inner parts of the MW, $\lesssim 30\,\mathrm{kpc}$, the track deviations, pole spreads and widths of streams have their largest fraction of perturbed streams. These statistics all show negative correlations with Galactocentric radius. However, the inner halo streams do not always show discernible differences for the choice of Galactic potential. Nevertheless, inner halo streams appear to be more affected by the response of the MW halo than by the LMC itself e.g., longitudinal pole spread.

\begin{figure*}
    \centering
    \includegraphics[width=\linewidth]{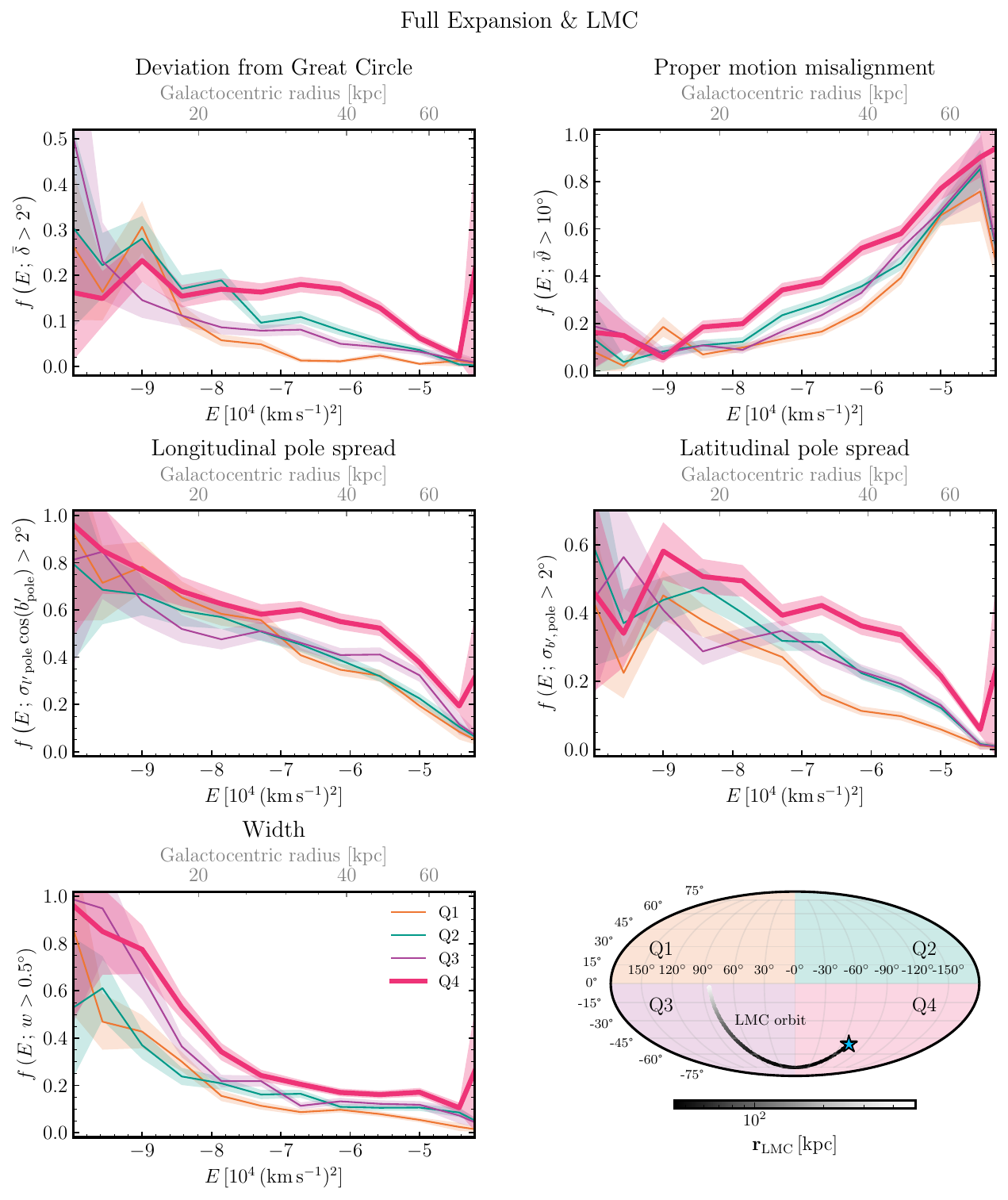}
    \caption{Same quantities as in Fig.~\ref{fig5}, now decomposed into quadrants in Galactic coordinates for the fiducial ‘Full Expansion \& LMC' potential. The colour scheme for each quadrant is reflected in the lower right panel which shows an all-sky projection in Galactic coordinates. The path of the LMC over the last 5 Gyrs is over-plotted, and coloured by its Galactocentric distance, with present-day as the blue star. Within the uncertainties, the direct effect of the LMC in Q4 (bold line) is distinguishable between the on-sky quadrants. This suggests that observations of streams across the whole sky will show a localised LMC effect on these quantities.}
    \label{fig6}
\end{figure*}

\subsubsection{Angular dependence}\label{sec:res-obs-effects-angdep}
To determine whether the LMC leaves a localised detectable signature, we investigate the properties of our stream population in the fiducial ‘Full Expansion \& LMC' potential by dividing the Galactic all-sky view into quadrants. We visualise these quadrants in the lower right panel of Fig.~\ref{fig6} with the past orbit of the LMC coloured by its Galactocentric distance.
We define Quadrant 1 (Q1, orange) as $l \in \{+180^{\circ}, 0^{\circ}\}$, $b \in \{0^{\circ}, +90^{\circ}\}$, Quadrant 2 (Q2, green) as $l \in \{0^{\circ}, -180^{\circ}\}$, $b \in \{0^{\circ}, +90^{\circ}\}$, Quadrant 3 (Q3, purple) as $l \in \{+180^{\circ}, 0^{\circ}\}$, $b \in \{-90^{\circ}, 0^{\circ}\}$ and  Quadrant 4 (Q4, pink) as $l \in \{0^{\circ}, -180^{\circ}\}$, $b \in \{-90^{\circ}, 0^{\circ}\}$. 

In Fig.~\ref{fig6} we show the same stream statistics as in Fig.~\ref{fig5} except now only for the fiducial ‘Full Expansion \& LMC' potential and split into these angular quadrants. For all of these statistics, Q4 (shown in bold) has the highest fraction of streams with detectable signatures of the LMC. At present-day the LMC resides in Q4 at a Galactocentric distance of $\sim50\,\mathrm{kpc}$ having just completed its pericentric passage. Around this distance, the detectable fraction of affected proper motion misalignments and pole spreads are significantly higher, $\sim30\%$, in Q4 compared to Q1 which is least affected.
Similarly, the track deviations and widths show this relative enhancement, albeit with a smaller amplitude, $\sim10 - 20\%$ differences between Q4 and Q1.
The relative enhancements between quadrants depend on properties of the adopted MW--LMC model. Hence, this provides a future opportunity to learn more about the MW--LMC system; see Sec.~\ref{sec:discussion-caveat}.
This angular dependence clearly demonstrates there is a local effect of LMC of MW streams. There is a direct relationship between proximity to the LMC and the likelihood of a stream being significantly perturbed. 

We replotted Fig.~\ref{fig6} to show the results for the ‘Full Expansion (no LMC)' potential (not shown here). In this case we find there is no angular dependence on the fraction of significantly perturbed streams for any of the stream statistics used, as might be expected for such a global perturbation via the MW halo response.
These results motivate using streams as observational probes in all-sky surveys as discussed in the next sections.

\section{Proper motion misalignments}\label{sec:res-pms}

To connect with observations, in Sec.~\ref{sec:res-pms-DES} we take observational data from DES \citep{Shipp2018, Shipp2019} for streams that have proper motion measurements and find the fraction of these observed streams that have detectable signatures in their proper motions; as defined in Sec.~\ref{sec:res-detectable-signatures}. We compare this against theoretical predictions for the misalignment angle given our simulated stream samples.
We then make a prediction for the upcoming Rubin LSST \citep{Ivezic2019} in  Sec.~\ref{sec:res-pms-lsst}.

\subsection{Comparison to DES}\label{sec:res-pms-DES}

We are able to compare our theoretical prediction for the fraction of streams with detectable signatures in their proper motions with stream data from DES. We are able to calculate the angular misalignment of the stream proper motion vectors \citep{Shipp2019} with respect to the stream track \citep{Shipp2018}. All of the DES streams reside in Q4, as shown in the right panel of Fig.~\ref{fig7}. The red points represent the Galactic coordinates of the DES streams.
In the left panel of Fig.~\ref{fig7}, we show the result for the mean distance of the DES streams as the red point. The distance uncertainty is determined by the standard deviation of the DES stream distance data. The fractional uncertainty is the binomial standard deviation.
Additionally, we show the Q4 result for the ‘Full Expansion \& LMC' (thick pink line), ‘Rigid Monopole \& LMC' (pink dotted line) and ‘Full Expansion (no LMC)' (pink dashed line) potentials.  
Within uncertainty, the DES result agrees with our prediction for Q4 using the fiducial ‘Full Expansion \& LMC' potential. However, this rudimentary measurement is subject to sample incompleteness. As the right panel of Fig.~\ref{fig7} demonstrates, this measurement is mainly local to the immediate area surrounding the LMC i.e. there is a large proportion of Q4 that has no currently observed streams. We tentatively expect the measured fraction of significantly affected streams to be lower and drop to be further in line with our prediction for Q4 as more of the quadrant is surveyed.
Finally, the ‘Rigid Monopole \& LMC' and ‘Full Expansion (no LMC)' potentials in Q4 predict much lower fractional values at the mean DES measurement distance, with $\sim 2\sigma$ significance. This further demonstrates the importance of the LMC in both causing the MW halo response and its direct gravitational influence, respectively. 

\subsection{Prediction for Rubin LSST}\label{sec:res-pms-lsst}

\begin{figure*}
    \centering
    \includegraphics[width=\linewidth]{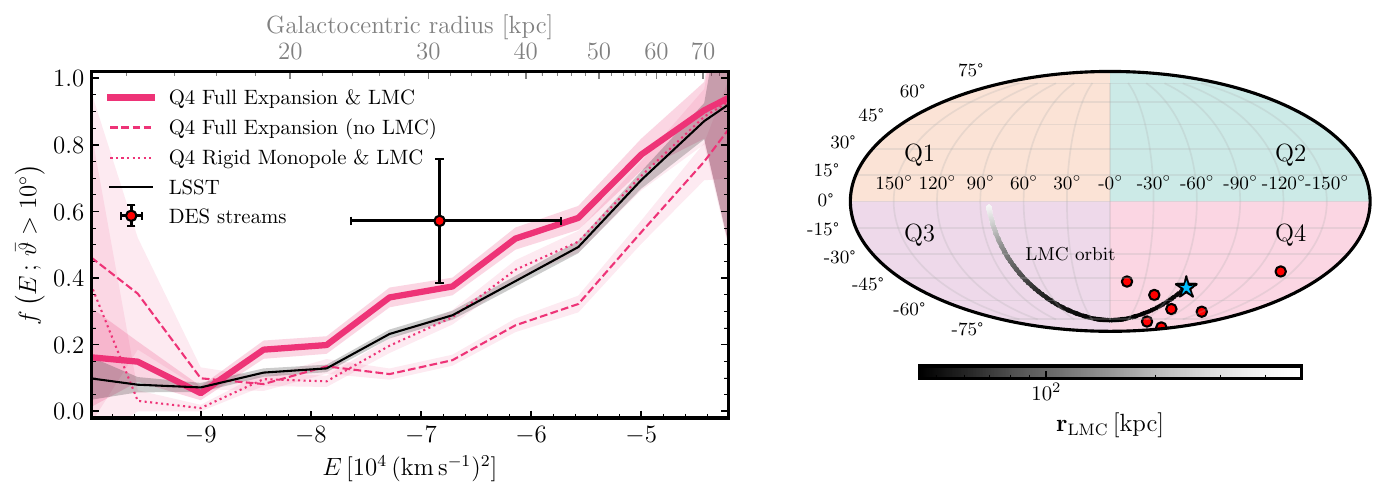}
    \caption{\textit{Left panel:} The fraction of streams with an ‘\textit{detectable}' signature of external perturbation in the misalignment of their proper motions with respect to the stream track as a function of the stream progenitor's energy (lower x-axis), or equivalently, radial Galactocentric position (upper x-axis). 
    The thick solid pink line represents the ‘Full Expansion \& LMC' potential in Q4 of the Galactic all-sky view; see the right panel. 
    The dotted/dashed pink lines show the result for Q4 using a MW--LMC potential described by the ‘Rigid Monopole \& LMC' / ‘Full Expansion (no LMC)' potential; Table.~\ref{table2}. 
    The black solid line represents the expected fraction for the footprint of the upcoming Rubin Observatory \textit{Legacy Survey of Space and Time}  \citep[LSST,][]{Ivezic2019}. 
    Poisson error bars are shown for each potential as the shaded region. 
    The fraction of streams with detectable signatures as measured using the \textit{Dark Energy Survey} \citep[DES,][]{Shipp2018, Shipp2019} streams with proper motion measurements is shown as the red circle with $1\sigma$ uncertainty.
    \textit{Right panel:} All-sky projection in Galactic coordinates split into quadrants. The path of the LMC over the last 2.5 Gyrs is over-plotted, and coloured by its Galactocentric distance, with the present-day as the blue star. The positions of the DES streams are shown as red circles. All of these streams are in ‘Q4' where the LMC is at present-day.}
    \label{fig7}
\end{figure*}

We are set to detect many more outer halo streams with the Rubin Observatory, in particular within the first 5 years with LSST \citep{Ivezic2019}. Using the fiducial ‘Full Expansion \& LMC' potential, we are able to forecast the expected fraction of streams with detectable signatures in their proper motions. To do this, we take the LSST sky footprint\footnote{\href{https://github.com/lsst/rubin_sim}{LSST GitHub repository:} Scheduler, survey strategy analysis, and other simulation tools for Rubin Observatory.}, which covers the entire southern sky, and find the fraction of simulated streams that fall within its boundaries. 
We show the prediction for the LSST footprint in the left panel of Fig.~\ref{fig7} as the solid black line. In comparison to the DES measurement, LSST will observe a lower fraction of significantly perturbed streams. This makes intuitive sense as LSST covers an area of the sky that is $\sim \times3-4$ larger, hence regions of higher and lower fraction of perturbed streams will be averaged out. The power of LSST will be to probe the outermost halo streams, $\gtrsim 40\,\mathrm{kpc}$, and determine their properties. This will allow us to better constrain the properties of the LMC as the properties of streams beyond this radial distance will be sensitive to both the mass ratio and infall trajectory of LMC; see Sec.~\ref{sec:discussion-caveat}. 
 
\section{Discussion}\label{sec:discussion}

\subsection{Reflex motion of the MW}\label{sec:discussion-reflex-motion}

The reflex motion in our Galaxy refers to the motion about the common center of mass in the MW--LMC system. This displaces stars and other objects in the inner and outer parts of our Galaxy in unique ways \citep{Weinberg1995, Gomez2015, Vasiliev2023}. We have considered two potentials which highlight the relative importance of this effect on inner/outer halo streams, namely the ‘Rigid MW without motion (no LMC)' and ‘Rigid MW + motion (no LMC)'. This reflex motion effect is seen in Fig.~\ref{fig5} for the longitudinal orbital pole spread and the width of streams. Within $\lesssim 25\,\mathrm{kpc}$, the inclusion of the reflex motion drives statistically significant differences between the fraction of streams with detectable perturbations, and grows in magnitude towards the innermost part of the halo.
Further, the probability density functions in Fig.~\ref{fig3} demonstrate how the reflex motion affects the distribution of stream statistics. It causes non-negligible effects on the orbital poles spreads and deviation from Great Circle orbits, however, these effects may not be ‘detectable' as the distributions do not have extended tails beyond the observational threshold values motivated in Sec.~\ref{sec:sumstats-dists}. It remains the scope of future studies to seek the effect of the reflex motion on the population of inner halo MW streams.

\subsection{Caveats}\label{sec:discussion-caveat}


The nature of our methodology poses a handful of caveats. The first is our use of only a single MW--LMC idealised simulation \citep{Lilleengen2023} which means we are unable to explore the dependence of our results on varying the masses of the MW, the LMC or both simultaneously. There is a handful of idealised MW--LMC simulations which adopt different choices of the MW and LMC masses \citep[e.g.,][]{Garavito-Camargo2019, Garavito-Camargo2021}. Using these simulations would be the next step in understanding the effect of halo mass, and infall trajectory, on MW stream perturbations. 
The perfect scenario would be to have many MW--LMC simulations, with varying masses of both haloes, to explore this effect on stellar streams and allow rapid inference in mass parameter space to find the ‘best-fit' masses. Cosmological simulations with analogue MW--LMC pairs offer an alternative route to study this \citep[e.g. the \textit{Milky Way-est} simulations,][]{Buch2024} although the infall times and orbital properties of the LMC analogue may not exactly match observational constraints. However, there remains considerable uncertainty in the orbital period and apocentric distance for the LMC \citep{Vasiliev2023}. Furthermore, although idealised MW--LMC simulations allow for easier control of model parameters and intuition of how they affect results, they cannot capture the triaxiality of dark matter haloes which cosmological simulations are able to. 

Another caveat comes from the choices of modelling the MW stream population. We choose an ergodic distribution function to draw the present-day positions and velocities of stream progenitors, Sec.~\ref{sec:methods-DFs}. Subsequently, this ensures that the generated stream distribution will be spatially isotropic. This approach will overlook any effects of clustering of streams i.e., around the LMC. Plus we do not model streams hosted by the LMC itself. Neglecting this population likely changes the predictions of angular on-sky dependencies for the fraction of significantly perturbed streams in Fig.~\ref{fig6} as Q4 would have a higher number of streams relative to other quadrants.
Further, we impose constraints on allowed orbital parameters; Sec.~\ref{sec:methods-DFs}. For a rigid MW potential, we would expect the proper motions of stream stars to be aligned with its track. However, in Fig.~\ref{fig5}, our rigid MW potential does not demonstrate this behaviour at larger Galactocentric distances. This can be explained by the fact that the misalignment is calculated in a Galactocentric and not a reflex-corrected Heliocentric reference frame. At these distances streams will be at, or close to, their apocenter, hence in the Galactocentric frame, their proper motions and stream track can be naturally misaligned. Therefore, the constraint on the allowed range of apocenters is degenerate with the expected fraction of streams with significantly misaligned proper motions. 

Finally, we did not include a dark matter subhalo population which may influence the dynamics and properties of the MW stream population. Modelling individual stream--subhalo interactions allows the recovery of the mass and density profile of the subhalo \citep{Erkal2015b, Bonaca2019, Hilmi2024}. Given the larger number of MW streams, a statistical approach would be to model a dark matter subhalo population and infer its cumulative effect on the population of streams \citep[e.g.,][see the latter for a novel Hamiltonian perturbation theory approach to modelling stream--subhalo perturbations in flexible potentials]{Carlberg2013, Arora2023, Nibauer2024}. Predominantly, dark subhalo impacts have been summarised by the power spectrum of density variations they imprint in streams \citep{Banik2021a, Banik2021b}, although this metric cannot always distinguish the cause of the density variation \citep[e.g.][]{Ibata2020}. Furthermore, the summary statistics used in this work to investigate LMC-induced effects may also be affected by the cumulative effect of many stream-subhalo interactions. We endeavour to add a realistic MW subhalo population to the fiducial MW--LMC potential used in this study. This would allow us to self-consistently generate and evolve the MW stream population in the presence of a dark matter subhalo population.

\subsection{Alternative dark matter models}\label{sec:discussion-alt-dm-models}

Often, MW--LMC simulations adopt collisionless cold dark matter to capture and study the deformations to the dark matter haloes of the MW and LMC \citep[e.g., those used in this work,][]{Lilleengen2023}. 
In fact, the structure of the deformations depends on the chosen dark matter model. Using self-interacting dark matter instead would subject both haloes to additional forces that depend on the relative velocities of both haloes throughout the interaction \citep[e.g.,][]{Furlanetto2002, Kaplinghat2016}.
Moreover, fuzzy dark matter (FDM) produces dynamically colder dark matter wakes with more granular structure \citep{Foote2023}. Plus, FDM naturally produces cored dark halo profiles which display different dynamical friction behaviour compared to cusped profiles \citep{Read2006}.
Furthermore, alternative gravity models, i.e., modified Newtonian Dynamics, do not have dark matter haloes to deform. Therefore, there is no effect from dynamical friction due to the MW dark halo \citep[e.g.,][]{Ciotti2004, Nipoti2008}. This would drastically alter the past orbit of the LMC \citep[e.g.,][]{Wu2008, Schee2013}, offering another route to constrain gravity and dark matter models.

\section{Summary and Conclusions}\label{sec:conclusions}

In this work, we have explored the properties of stellar streams in time-dependent MW--LMC systems. For the first time, we have taken the approach to model a realistic stellar stream population in deforming MW--LMC systems. This has allowed us to statistically determine the significance of dynamical signatures imprinted into a set of stream summary statistics used to describe each stream. The statistics used are the stream length, asymmetry, width, deviation from Great Circle orbits, local velocity dispersion, proper motion misalignment, and orbital pole dispersions. 
We investigated the radial and on-sky angular dependence of stream perturbations caused by the direct effect of stream–LMC interactions and/or the response of the MW dark matter halo.
Our use of BFEs to describe the MW--LMC system allows us to flexibly alter the potential to investigate these effects. Our main conclusions are:
\begin{enumerate}
    \item For outer halo streams, the width, deviation from a Great Circle, proper motion misalignment, and spread in orbital pole stream statistics are \textit{dependent} on the choice of Galactic potential used to generate the MW stream population, particularly for those with and without an LMC; see Sec.~\ref{sec:sumstats-dists} \& Fig.~\ref{fig3}.
    \item The direct effect of the LMC on streams is seen to increase the fraction of outer halo streams with ‘detectable' signatures of an external perturbation by up to e.g., $\sim25\%$ in significantly misaligned proper motions compared to a potential without an LMC; see Sec.~\ref{sec:res-obs-effects-radialdep} \& Fig.~\ref{fig5}. Further, we find there is an on-sky angular dependence with the highest fractions for all statistics coinciding with the quadrant of the sky where the LMC is located at present-day; see Sec.~\ref{sec:res-obs-effects-angdep} \& Fig.~\ref{fig6}.
    \item The effect of the MW halo response is dominated by its dipole harmonic. This effect leaves a signature in the longitudinal and latitudinal pole spreads, distinct from the direct LMC influence, for streams at distances $\lesssim 25\,\mathrm{kpc}$. While the dipole and LMC effects are distinct, it is inconclusive whether this will be detectable. For a MW halo that is initially spherically symmetric, the monopole and quadrupole harmonics have little, if any, effect on further dynamically altering stream statistics; see Sec.~\ref{sec:res-obs-effects-radialdep} \& Fig.~\ref{fig5}.
    \item For outer halo streams, the length, asymmetry and local velocity dispersion stream statistics show hardly any dependence of the choice of Galactic potential used to generate the MW stream population; see Sec.~\ref{sec:sumstats-dists} \& Fig.~\ref{fig3}.
    \item We compare our prediction for the fraction of streams with significantly misaligned proper motions to the observational data from DES. Within uncertainty, our prediction agrees with the DES data despite an incomplete data sample; see Sec.~\ref{sec:res-pms-DES} \& Fig.~\ref{fig7}.
    \item We make a prediction for the fraction of streams with significantly misaligned proper motions as a function of distance for the upcoming Rubin LSST survey. With more LSST data for outer halo streams, this will allow us to improve constraints on the properties of the LMC e.g. its mass and infall trajectory; see Sec.~\ref{sec:res-pms-lsst} \& Fig.~\ref{fig7}. 
\end{enumerate}

Our results have demonstrated that MW stellar streams are affected both directly by interactions with the LMC and by the response of the MW halo to the infalling satellite. The extent to which features of streams are affected is dependent on its radial and angular location.
We find that we can, and already have, detected the direct influence of the LMC on streams within its immediate vicinity. 
All-sky surveys to find outer halo streams are increasingly important given the relative enhancement in the fraction of streams with ‘detectable' signatures across on-sky quadrants.
The reflex motion, captured by the dipole of the MW halo response, appears to mostly affect streams in the inner MW halo. This motivates future studies committed to investigating the innermost streams in our Galaxy.
With the advent of more stream data from Rubin LSST, we look forward to comparing our theoretical predictions with the data to better understand the MW--LMC interaction, properties of dark matter and the LMC itself.
\newline

\section{Data availability}\label{sec:data}

Greek mythology understands \textit{Oceanus}\footnote{\url{https://en.wikipedia.org/wiki/Oceanus}}, Son of Gaia, to be the primordial Greek titan god of the great, earth-encircling stream $\Omega\kappa\epsilon\alpha\nu o\varsigma$, and, amongst many other roles, the font of all fresh-water, e.g rain-clouds. 
Hence we name our populations of simulated stellar streams generated after Oceanus. 
We publish the $t = 0\,\mathrm{Gyr}$ simulation snapshots of all $16,384$ streams in each MW--LMC potential used in this work. Each stream contains $20,000$ particles with 3-D Galactocentric positions and velocities, plus progenitor information on its initial conditions, mass and scale radius. We do not include summary statistic values. The functions to calculate stream summary statistics will be shared upon reasonable request. This dataset can be used to explore the effect of the LMC and the MW halo response on simulated MW streams. The archive with this data is hosted on Zenodo:~\dataset[\textit{oceanus}:~Populations of stellar streams in deforming MW--LMC systems - I]{https://doi.org/10.5281/zenodo.13771517} \& \dataset[II]{https://doi.org/10.5281/zenodo.13771837}.
The python interface to integrate orbits and access to the expansion model for the \citet{Lilleengen2023} MW--LMC simulation can be found here: \url{https://github.com/sophialilleengen/mwlmc}.

\section{Acknowledgments}

RANB acknowledges support from the Royal Society and the Flatiron Institute pre-doctoral program, funded by the Simons Foundation. RANB would like to thank Julianne Dalcanton, David Hogg, Danny Horta, and all of the Nearby Universe group at the Center for Computational Astrophysics for insightful discussions. JLS acknowledges support from the Royal Society (URF\textbackslash R1\textbackslash191555). NGC would like to thank Peter Ferguson for pointing us toward how to generate the LSST footprint. 



\software{Astropy \citep{astropy:2013, astropy:2018, astropy:2022}, NumPy \citep{harris2020array}, SciPy \citep{2020SciPy-NMeth}, Matplotlib \citep{Hunter:2007}, Eigen \citep{eigenweb}, \textsc{exp} \citep{2022MNRAS.510.6201P}, \textsc{mwlmc} \citep{Lilleengen2023}, \textsc{agama} \citep{2019MNRAS.482.1525V}, \textsc{gala} \citep{gala} and \href{https://github.com/lsst/rubin_sim}{LSST GitHub repository}.}



\appendix

\section{Frames of reference and fictitious forces}\label{app:A}

\begin{figure}
    \centering
    \includegraphics[width=\linewidth]{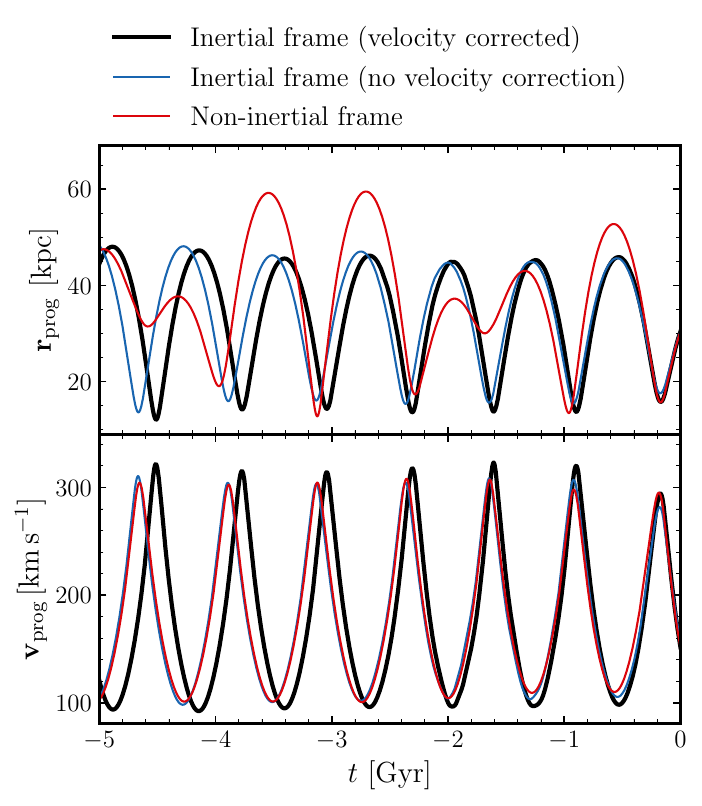}
    \caption{Example progenitor orbit when evaluating the forces for leapfrog integration in the frame of reference of the velocity corrected inertial (thick black), the non-velocity corrected inertial (blue) and the non-inertial (red). The upper panel shows the radial positions, while the lower panel shows the magnitude of the velocity vector. A reproducible version of this figure is available via this \href{https://gist.github.com/dc-broo3/3ea3e8b15ebc2ad57d31f06cca5fbf75}{GitHub Gist}.}
    \label{fig8}
\end{figure}

In this section, we build upon Sec.~\ref{sec:methods-dynmodels} to describe the process we use to generate a stellar stream. The first step to generate a stream is to integrate the progenitor's orbit backwards through the desired integration time. To achieve this, one must start from the present-day Galactocentric initial conditions drawn from the tracer Dehnen distribution function \citep{Dehnen1993}. To ensure we integrate in the inertial frame of reference, we must account for the current displacement of the Galactic barycenter due to the merger with the LMC. To do this, we add the positions and velocities of the barycenter to the present-day Galactocentric phase-space coordinates i.e., $\mathbf{r}_{\mathrm{0,\,inertial}} = \mathbf{r}_{0,\,galactocentric} + \mathbf{r}_{\mathrm{barycenter}}$ and $\mathbf{v}_{0,\,\mathrm{inertial}} = \mathbf{v}_{0,\,galactocentric} + \mathbf{v}_{\mathrm{barycenter}}$. We integrate the progenitor between successive time steps using the leapfrog integrator\footnote{\url{https://en.wikipedia.org/wiki/Leapfrog_integration}} implemented in \textsc{gala} \citep{gala}. We adopt a constant time step of $2\,\mathrm{Myr}$. We have checked this integrator against our own leapfrog implementation and recover the same orbital trajectory. The force evaluations at each time step are calculated using the \textsc{mwlmc} simulation of \citet{Lilleengen2023}. We use the functions that return forces in the frame of reference of each individual component, e.g.~the ‘\textit{mwhalo\_fields}' function returns the forces of the MW halo in its own reference frame. To shift each force evaluation to the inertial frame, we subtract the current expansion centre for each component before evaluating the force using the ‘\textit{expansion\_centres}' function at each time, e.g., $\mathbf{a}_{\mathrm{halo,\,inertial}}(\mathbf{r} - \mathbf{r}_{\mathrm{halo\, center}})$. When the progenitor has been integrated backwards through the desired time, we subtract the barycenter displacement and motion at this time to return the initial conditions in the Galactocentric Cartesian frame, i.e., $\mathbf{r}_{\mathrm{f,\,galactocentric}} = \mathbf{r}_{f} - \mathbf{r}_{\mathrm{barycenter}}$ and $\mathbf{v}_{f,\,\mathrm{galactocentric}} = \mathbf{v}_{f} - \mathbf{v}_{\mathrm{barycenter}}$.

To generate the stellar stream, we begin from the cluster's initial conditions in the Galactocentric Cartesian frame. Again, to ensure forces are evaluated in the inertial frame we first add the barycentric displacement and motion at the start time. The forward integration to generate the stream by the mLCS technique (Sec.~\ref{sec:methods-dynmodels}) with the forces evaluated in the inertial frame of reference.  At the final saved output, we subtract the barycentric displacement and motion to return the stream positions and velocities in the Galactocentric Cartesian frame.

The generation of a stellar stream must use the forces of the Galactic potential evaluated in the inertial frame of reference, otherwise, the stream will be subject to fictitious, non-physical, forces. Fictitious forces are pseudo-forces resulting from the frame of reference itself being accelerated or undergoing rotational motion. For example, if the forces are taken from the reference frame centered on the Milky Way disc, this will include fictitious forces. If one integrates orbits using forces evaluated in a non-inertial frame of reference, the resulting trajectory will be biased by these forces which we would not see in observation. To illustrate this in Fig.~\ref{fig8}, we integrate an example progenitor (‘\textit{stream\_0}' of our present-day globular cluster sample) in a MW--LMC potential described by the full basis function expansions, see Sec.~\ref{sec:methods-galpot}. We use the integration forces by evaluating them in the velocity-corrected inertial (thick black), the non-velocity corrected inertial (blue), and the non-inertial (red) frames of reference. We show the progenitor's radial position and velocity magnitude throughout its orbit in the upper and lower panels of Fig.~\ref{fig8}, respectively. By not accounting for the motion of the MW barycenter, and/or evaluating forces in a non-inertial frame of reference, the orbit of the progenitor varies drastically. The fictitious forces are apparent when comparing the non-inertial frame to the velocity-correct inertial one, causing the past orbit to become chaotic. 

\section{Distribution of statistics within streams}\label{app:B}

\begin{figure*}
    \centering
    \includegraphics[width=1\linewidth]{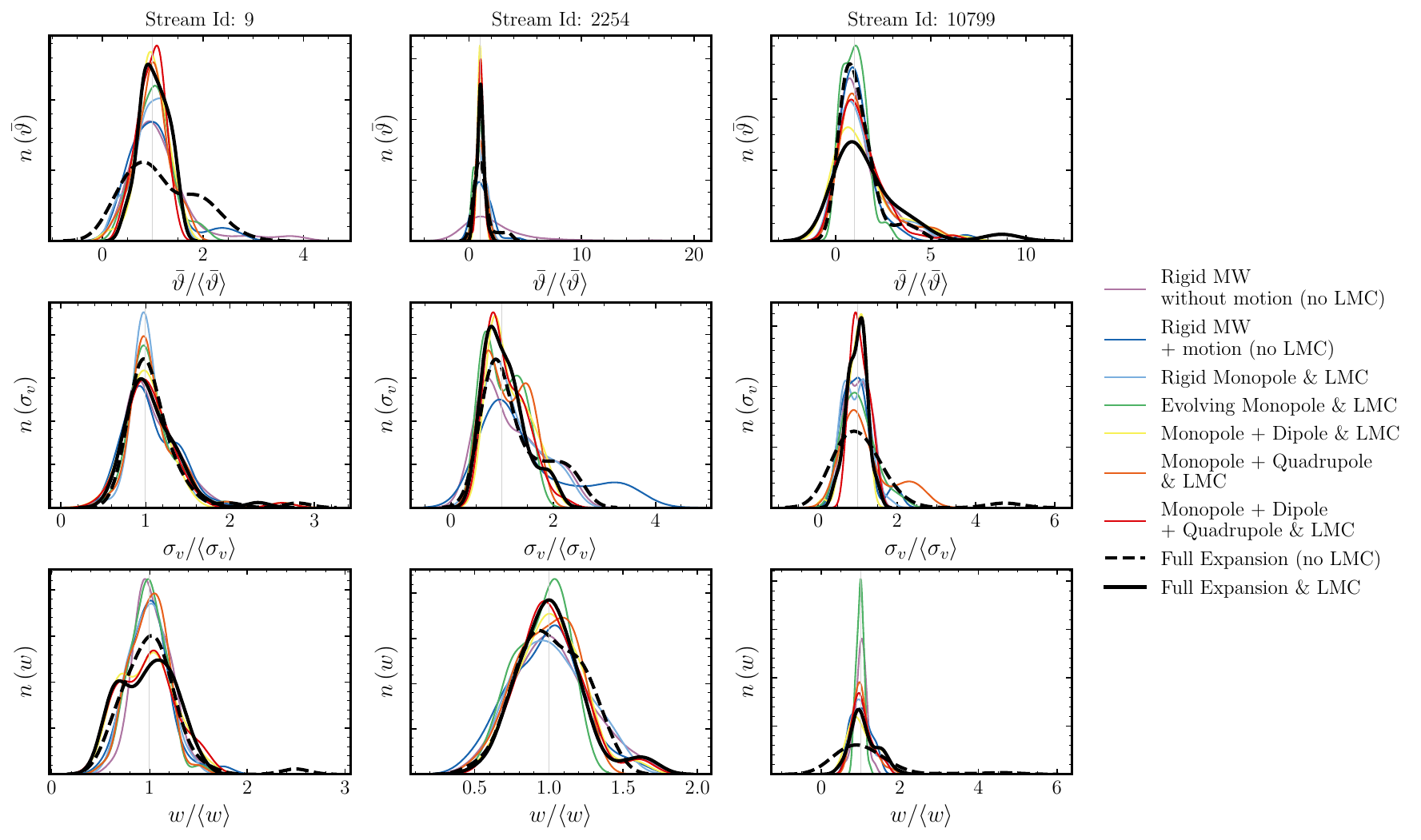}
    \caption{The distributions of the stream statistics computed in $50$ equally-spaced bins along three randomly-selected streams for the different potential setups (see Table.~\ref{table2}). For each stream and potential, the distribution is normalised by the median stream statistic over all bins (as used throughout this study) and shown using a kernel density estimate.
    In rows top to bottom, we show the distributions for the stream proper motion misalignments, local velocity dispersions and widths.
    Each column represents a unique stream from the full simulated population with the stream Id as the column title; see Sec.~\ref{sec:methods-gcpops}.
    In most potentials, the distributions are not significantly skewed. This implies that using other choices for the stream statistic would not qualitatively change the results in the main body.}
    \label{fig9}
\end{figure*}

In the main body, we have summarised the impact of perturbations on streams by using the median of statistics averaged over each stream. However, some streams may only be impacted along part of their length so this choice may `wash out' some signal.
To justify our choice of dian statistic across the entire stream, we make a comparison to the stream statistic value calculated in each bin across the streams; see Sec.~\ref{sec:sumstats}.
In Fig.~\ref{fig9}, we show the distributions of the statistics computed in each bin along the stream normalised by the median evaluated over all bins. For three randomly chosen streams from the populations in each potential, we show the distributions for the stream proper motion misalignments, local velocity dispersions and widths.
In most potentials, the distributions are not significantly skewed showing that all bins in each stream are similarly affected. Furthermore, the normalised statistic distributions for each potential setup are similar suggesting further statistics would not yield more discriminating power. This implies that, on average across our mock sample of streams, using other choices for the stream statistic would not qualitatively change the results. 
\citet{Erkal2019} show in their fig.~1 that the Orphan-Chenab stream is $\sim 200^{\circ}$ long and report proper motion misalignments in $\sim25^{\circ}$ bins. In our Fig.~\ref{fig4}, we show that the median length of simulated streams in all potentials is $\sim25^{\circ}$. Hence, on average, our streams are the same angular length as a single bin of the Orphan-Chenab stream where definitive proper motion misalignments are measured. Physically, this suggests that using median statistics to capture the behaviour across an entire stream in our simulated samples is reasonable given recent observations. For streams that are significantly longer than $\sim25^{\circ}$ this may not be fully representative.

\bibliography{manuscript}{}
\bibliographystyle{aasjournal}



\end{document}